\documentclass[aps,twocolumn,nofootinbib,superscriptaddress,floatfix]{revtex4}

\usepackage{color,amsmath,amssymb,graphicx,latexsym,subfigure}
\usepackage{threeparttable}
\usepackage[colorlinks]{hyperref}
\usepackage{ulem,soul}
\def\nat{Nature\ }
\def\aap{Astron.\ Astrophys.\ }
\def\apj{Astrophys.\ J.\ }
\def\apjl{Astrophys.\ J.\ Lett.\ }
\def\apjs{Astrophys.\ J.\ Supp.\ }

\def\mnras{Mon.\ Not.\ Roy.\ Astron.\ Soc.\ }

\def\prd{Phys.\ Rev.\ D\ }
\def\prl{Phys.\ Rev.\ Lett.\ }

\def\araa{Annu.\ Rev.\ Astron.\ Astrophys.\ }

\begin{document}

\title{Rapidly growing primordial black holes as seeds of the massive high-redshift JWST Galaxies}

\author{Guan-Wen Yuan}
\affiliation{Key Laboratory of Dark Matter and Space Astronomy, Purple Mountain Observatory, Chinese Academy of Sciences, Nanjing 210023, People's Republic of China}
\affiliation{School of Astronomy and Space Science, University of Science and Technology of China, Hefei 230026, People's Republic of China}
\affiliation{Institute of High Energy Physics, Chinese Academy of Sciences, Beijing 100049, China}

\author{Lei Lei}
\affiliation{Key Laboratory of Dark Matter and Space Astronomy, Purple Mountain Observatory, Chinese Academy of Sciences, Nanjing 210023, People's Republic of China}
\affiliation{School of Astronomy and Space Science, University of Science and Technology of China, Hefei 230026, People's Republic of China}

\author{Yuan-Zhu Wang}
\affiliation{Key Laboratory of Dark Matter and Space Astronomy, Purple Mountain Observatory, Chinese Academy of Sciences, Nanjing 210023, People's Republic of China}

\author{Bo Wang} 
\affiliation{School of Astronomy and Space Science, University of Science and Technology of China, Hefei 230026, People's Republic of China}
\affiliation{Deep Space Exploration Laboratory/Department of Astronomy, School of Physical Sciences, University of Science and Technology of China, Hefei 230026, People's Republic of China}

\author{Yi-Ying Wang}
\affiliation{Key Laboratory of Dark Matter and Space Astronomy, Purple Mountain Observatory, Chinese Academy of Sciences, Nanjing 210023, People's Republic of China}
\affiliation{School of Astronomy and Space Science, University of Science and Technology of China, Hefei 230026, People's Republic of China}

\author{Chao Chen} 
\affiliation{Jockey Club Institute for Advanced Study, The Hong Kong University of Science and Technology, Hong Kong, China}

\author{Zhao-Qiang Shen}\email{zqshen@pmo.ac.cn}
\affiliation{Key Laboratory of Dark Matter and Space Astronomy, Purple Mountain Observatory, Chinese Academy of Sciences, Nanjing 210023, People's Republic of China}

\author{Yi-Fu Cai}\email{yifucai@ustc.edu.cn}
\affiliation{School of Astronomy and Space Science, University of Science and Technology of China, Hefei 230026, People's Republic of China}
\affiliation{Deep Space Exploration Laboratory/Department of Astronomy, School of Physical Sciences, University of Science and Technology of China, Hefei 230026, People's Republic of China}

\author{Yi-Zhong Fan}\email{yzfan@pmo.ac.cn}
\affiliation{Key Laboratory of Dark Matter and Space Astronomy, Purple Mountain Observatory, Chinese Academy of Sciences, Nanjing 210023, People's Republic of China}
\affiliation{School of Astronomy and Space Science, University of Science and Technology of China, Hefei 230026, People's Republic of China}

\begin{abstract}

A group of massive galaxies at redshifts of $z\gtrsim 7$ have been recently detected by the James Webb Space Telescope (JWST), which were unexpected to form so early within the framework of standard Big Bang cosmology. In this work, we propose that this puzzle can be explained by the presence of some primordial black holes (PBHs) with a mass of $\sim 1000 M_\odot$. These PBHs, clothed in dark matter halo and undergoing super-Eddington accretion, serve as seeds for the early galaxy formation with masses of $\sim 10^{8}-10^{10}~M_\odot$ at high redshift, thus accounting for the JWST observations. Using a hierarchical Bayesian inference framework to constrain the PBH mass distribution models, we find that the Lognormal model with $M_{\rm c}\sim 750M_\odot$ is preferred over other hypotheses. These rapidly growing BHs are expected to emit strong radiation and may appear as high-redshift compact objects, similar to those recently discovered by JWST. Although we focuse on PBHs in this work, the bound on the initial mass of the seed black holes remains robust even if they were formed through astrophysical channels.

\end{abstract}

\date{\today}

\maketitle
 
\section{Introduction}

Understanding the beginning of galaxy formation at the end of the cosmic dark ages, where massive black holes (MBHs) are generally thought to reside, is a key goal of modern cosmology~\cite{Barkana:2000fd, 2011ARA&A..49..373B}. One fascinating possibility is that MBHs in high-redshift galaxies could be seeded by primordial black holes (PBHs)~\cite{ Volonteri:2021sfo, Clesse:2015wea}. Consequently, the investigation on PBHs offers an exciting opportunity to explore the mechanics of the early Universe~\cite{Escriva:2022duf}. 

The presence of PBHs was originally proposed by Zel'dovich and Hawking~\cite{1966AZh, 1971MNRAS}, and it was speculated that these exotic objects could form through the gravitational collapse of overdense regions in the early Universe. Given the increasingly stringent null results from searches for particle dark matter~\cite{Bertone:2016nfn, Schumann:2019eaa} (see however \cite{2022PhRvL.129i1802F}), the possibility of PBHs as a class of DM candidates has received wider and wider attention~\cite{1975Natur.253..251C, Carr:2016drx}. While various observations over several mass ranges have constrained the abundance of PBHs as DM candidates (for comprehensive review, see Ref.~\cite{Carr:2020gox}), in particular, the massive PBHs $(10 M_{\odot} \lesssim M \lesssim 10^4 M_{\odot})$ may account for only a small fraction of the DM. Nonetheless, they could still be responsible for the MBHs observed at high redshift ($z \gtrsim 7$)~\cite{Inayoshi:2019fun}. 

%%% the possibility of PBH seeding SMBH
Fortunately, the ongoing JWST observations are expected to shed further light on the nature of high-redshift galaxies and the role of PBHs in their formation~\cite{Gardner:2006ky,2019ApJ...877...23L, Volonteri:2021sfo, Massonneau:2022sye}. Recent JWST observations have identified several bright galaxy candidates at $z \gtrsim 7$~\cite{2022arXiv220712446L, 2023MNRAS.519.1201A, 2023ApJ...942L...9Y, 2023MNRAS.518.2511L}. One possible interpretation is that PBHs have altered the matter power spectrum~\cite{Liu:2022bvr, 2023ApJ...944..113B, 2023MNRAS.519.4753T, 2023PhRvD.107d3502H, 2023arXiv230109403C, Wang:2023len}. 
In this work, we show that PBHs with masses of approximately $1000 M_{\odot}$ could potentially grow up into MBHs with masses of $10^{4}-10^{8} M_{\odot}$ through super-Eddington accretion~\cite{Wang:2013ha} within the dark matter halo. We then use a hierarchical Bayesian inference framework to constrain the PBH mass distribution models, and we find that the Lognormal model with the $M_{\rm c}\sim 750M_\odot$ is strongly preferred over other hypotheses. These growing PBHs, with their strong radiation, may appear as high-redshift compact objects, accounting for the observations made by JWST.

\section{PBH Evolution in the Early Universe}
After having been generated, PBHs may evolve through accretion throughout cosmic history, and this accretion of baryonic matter onto PBHs can significantly impact their masses~\cite{DeLuca:2020fpg, Serpico:2020ehh,Hasinger:2020ptw}. However, the physics of accretion is complex, as the accretion rate and the geometry of the accretion flow are intertwined and both play a crucial role in the evolution of the PBH masses. An analytic accretion model, such as the one developed by Mack, Ricotti and Ostriker~\cite{Mack:2006gz,Ricotti:2007jk,Ricotti:2007au}, can be used to study the accretion of baryonic matter onto PBHs and provide useful insights into the accretion behaviour. 
In the intergalactic medium, a PBH with mass $M$ can accrete baryonic matter at the Bondi-Hoyle rate $\dot{M}_{\mathrm{B}}$, which can be expressed as
\begin{equation}\label{eq::MdotB}
\dot{M}_{\mathrm{B}}=4 \pi \lambda m_{H} n_{\mathrm{gas}} v_{\mathrm{eff}} r_{\mathrm{B}}^{2} ,     
\end{equation}
where $r_B = GM/v_{\rm eff}^2$ is the Bondi-Hoyle radius, $n_{\rm gas}$ is the hydrogen gas number density, and $v_{\rm eff}=\sqrt{v_{\rm rel}^2 + c_s^2}$ is the PBH effective velocity, which is defined in terms of the PBH relative velocity $v_{\rm rel}$ and the gas with sound speed $c_s$. The accretion parameter $\lambda$ accounts for the gas viscosity, the Hubble expansion, and the Compton scattering between the gas and the CMB, defined in Eq.(23) in~\cite{Ricotti:2007au}.

It is important to note that since PBHs constitute only a tiny fraction of the dark matter (DM) in the Universe, one must also consider the dominant dark matter component forming a dark halo around each PBH~\cite{Ricotti:2007au, DeLuca:2020fpg}. Simulations~\cite{1985ApJS...58...39B, Ricotti:2007au} suggest that this dark halo has a density profile $\rho \sim r^{-2.25}$ and a  mass given by
\begin{equation}
M_h(z)=3M_{\rm BH} \left(\frac{1+z}{1000} \right)^{-1},
\end{equation}
which increases over time, provided the PBH does not interact with others. Due to the weak interaction between dark matter particles and between dark matter and baryonic matter, the timescale for dark matter particles to lose their angular momentum and fall into the PBH is much longer than that for baryonic matter. Consequently, the direct accretion of dark matter in the accretion model is negligible~\cite{Ricotti:2007au}.

As a result, more and more dark matter particles are attracted by the PBH-dark matter halo system over time, causing the dark matter halo to grow larger and become more spherical. On the other hand, the effect of this dark matter halo, or ``dark matter clothing", is to enhance the gas accretion rate, acting as a catalyst. More specifically, if the Bondi radius $r_B$ of PBH is greater than twice the characteristic halo radius $r_h$, the gas accretion rate onto the system is the same as that onto a bare PBH with a dark halo mass $M_h$. However, if $r_B$ is smaller or $r_h$ is larger, the accretion enhancement caused by the halo must be accounted for, requiring modifications to the accretion parameter as detailed in~\cite{Ricotti:2007jk} and~\cite{Ricotti:2007au}.

However, the gas accretion rate will be suppressed by the outflows from the PBH which sweep away the surrounded medium and only leave the diluted and hot gas.
The integration of this effect, known as mechanical feedback (MF), into the analytic accretion calculation poses significant challenges~\cite{Ali-Haimoud:2016mbv, DeLuca:2020bjf}. However, a glimmer of optimism arises from recent 3D numerical simulations~\cite{Bosch-Ramon:2020pcz, Bosch-Ramon:2022eiy}. These simulations have provided valuable insights, suggesting a roughly consistent fractional rescaling factor for PBH accretion. As a result, we can succinctly express PBH accretion as $\dot{M}_{\mathrm{acc}}= f_{\rm MF} \dot{M}_{\mathrm{B}}$, with an adopted value of $f_{\rm MF} \simeq 0.1$ as Ref~\cite{Piga:2022ysp}. This selection is derived from averaging the potential outflow orientation angles during the accretion process. While this reduction in the accretion rate is significant, it serves as an illustrative example of PBH evolution in the early universe. A comprehensive analysis of these additional phenomena is a subject for future investigation, expanding our understanding of this intricate process.

After considering the effect of the dark halo and the feedback, the efficiency and shape of the accretion process are entirely encoded in the dimensionless baryonic accretion rate $\dot{m}$, which is defined as  $\dot{m} = \dot{M}_{\rm acc} / \dot{M}_{\rm Edd}$ with the Eddington accretion $\dot{M}_{\rm Edd} = 1.44\times 10^{17} (M/M_{\odot})\,{\rm g\,s^{-1}}$.

\begin{figure}[htbp] 
\centering 
\includegraphics[width=0.96\linewidth]{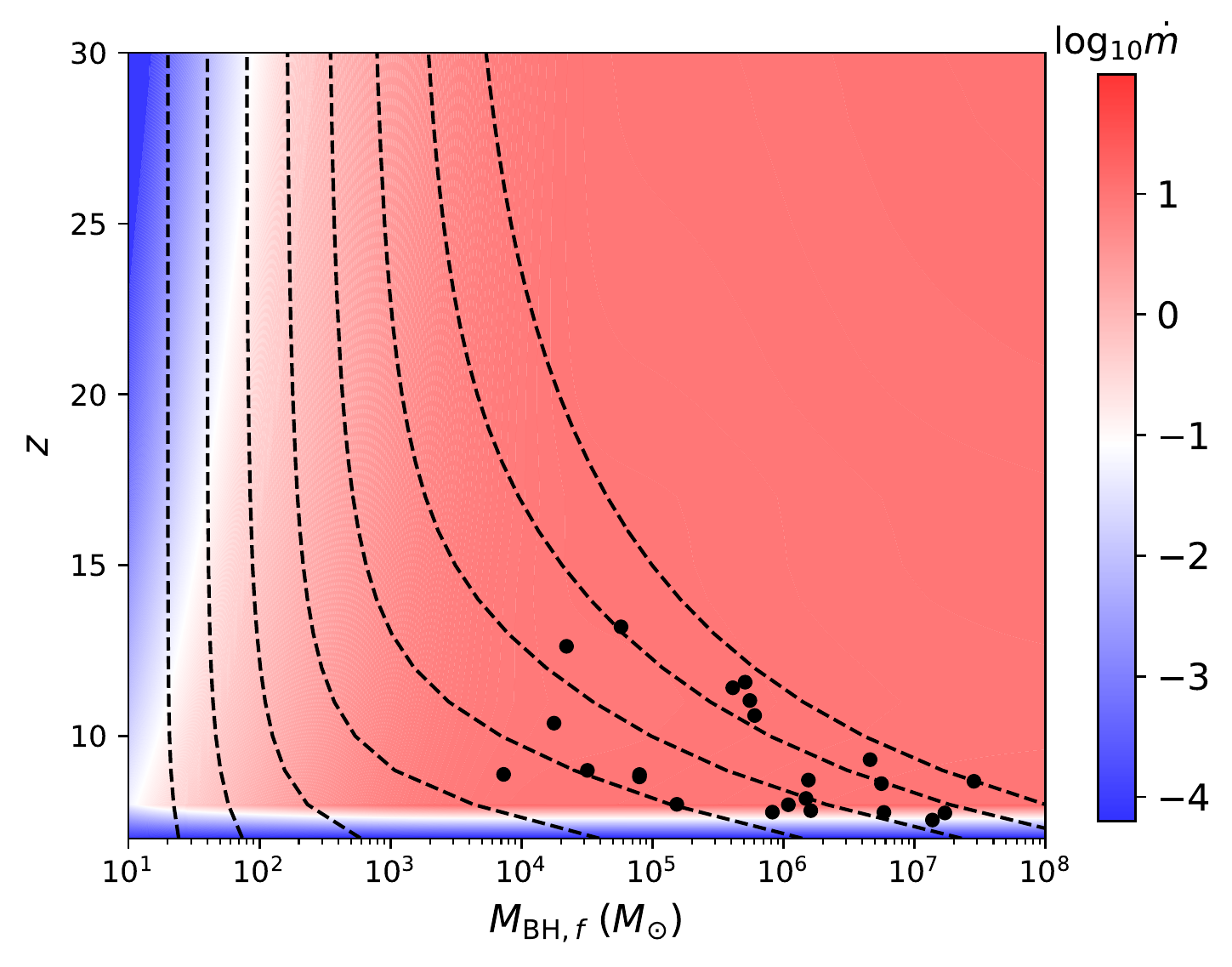} 
\caption{The accretion rate parameter $\dot{m}$ is shown as a function of the final mass of PBH $M_{{\rm BH}, f}$ and redshift $z$. After considering the growth of surrounding dark matter halo and the effects of feedback mechanism, seed black hole in the early universe could achieve stable super-Eddington accretion. The black points in the figure represent galaxy candidates identified from JWST observations, with a relation of $M_{\star} - M_{\rm BH}$. The figure also shows trajectories of individual PBHs with $20M_{\odot}$, $40M_{\odot}$, $80M_{\odot}$, $160M_{\odot}$, $320M_{\odot}$, $640M_{\odot}$, $1280M_{\odot}$, and $2560M_{\odot}$ in the $(M, z)$ plane, represented by black dotted lines. } 
\label{pbh_mass_evolution}
\end{figure}

In Figure~\ref{pbh_mass_evolution}, we present the accretion rate $\dot{m}$ as a function of PBH mass and redshift, utilizing data sources listed in Table~\ref{data_table} and the relationship between $M_{\star}$ and $M_{\rm BH}$. The figure illustrates that in the early universe, seed black holes surrounded by dark matter halo can keep stable super-Eddington accretion because the rapidly growing halo provide gravitational potential and enhance accretion.
The black dotted lines represent the trajectories of PBH evolution for various initial masses, including $20M_{\odot}$, $40M_{\odot}$, $80M_{\odot}$, $160M_{\odot}$, $320M_{\odot}$, $640M_{\odot}$, $1280M_{\odot}$ and $2560M_{\odot}$ within the $(M, z)$ plane. It is worth noting that the redshift cutoff for PBH accretion is set at 7.5 due to significant uncertainties related to large-scale structure. Typically, the PBHs with initial masses, denoted as $M_{\rm BH, i}$, less than a few solar masses face challenges in achieving efficient growth due to their low accretion rates. However, for PBHs with higher initial masses, the accretion process can be significantly more effective. Some PBHs that originated in the very early Universe have the potential to grow to a size approaching $\sim 10^{7}~M_\odot$~\cite{2022PhRvD.106d3539K}.

\begin{figure}[htbp] 
\centering 
\includegraphics[width=0.49\textwidth]{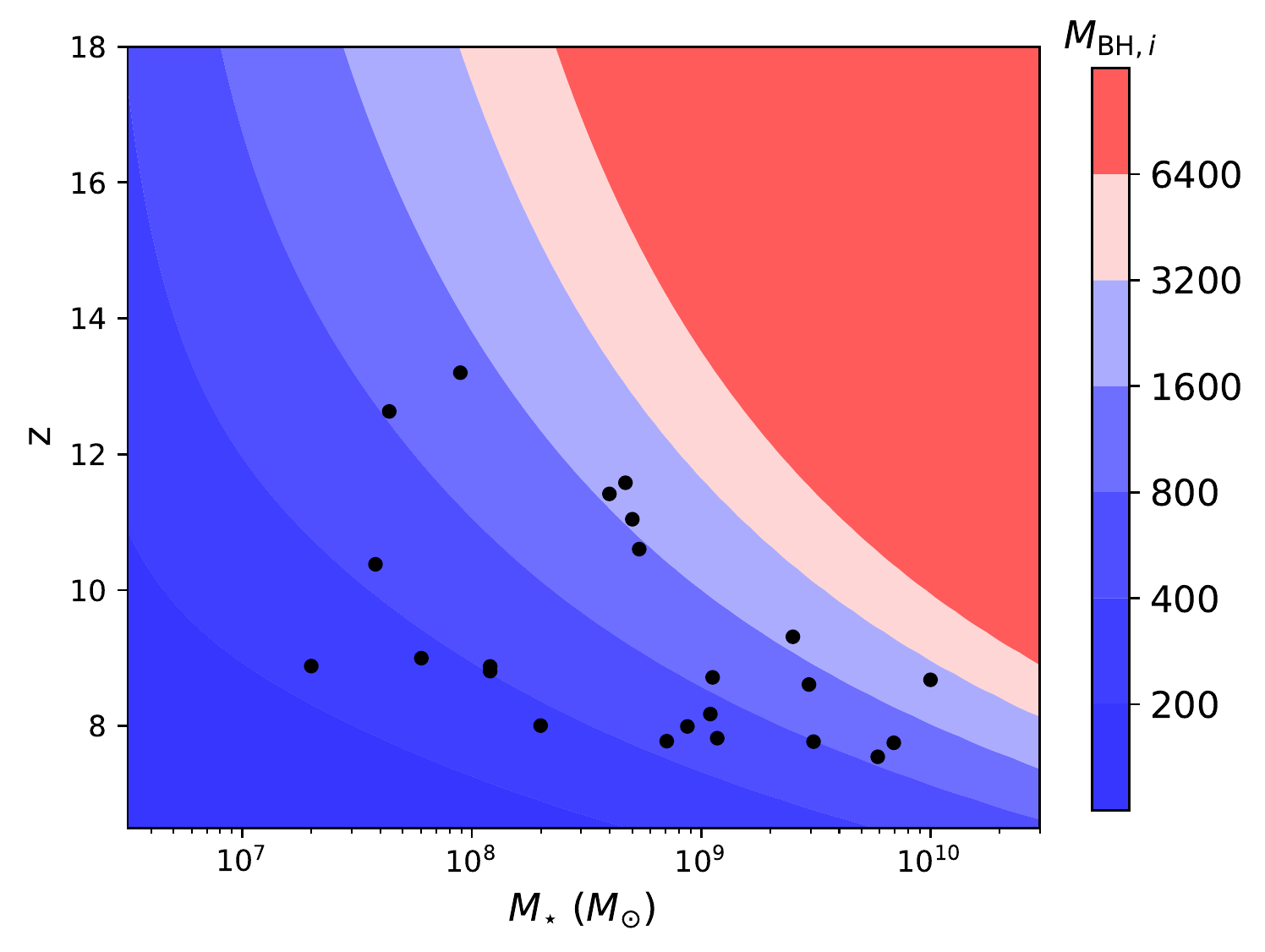} 
\caption{Initial PBH mass ($M_{\rm BH, i}$, represented by the color bar) as a function of the galaxy mass $M_{\star}$ and the redshift $z$. The candidates observed by JWST are represented by the black points, which are listed in Table~\ref{data_table}.} 
\label{contourf}
\end{figure}

In the local Universe, a well-established trend reveals a shared growth pattern between MBHs and galaxies, a relationship extensively discussed in the review by~\cite{Greene:2019vlv}. This co-evolution scenario is believed to be influenced by feedback processes originating from active galactic nuclei (AGNs) and the initial mass density of MBHs and galaxies. An empirical relationship between the mass of the MBHs, denoted as $M_{\rm BH}$, and the galaxy mass $M_{\star}$ is described by ${\rm log} (M_{\rm BH}/{M_{\odot}})  =\alpha+\beta {\rm log}(M_{\star}/{M_0})+\epsilon {\rm log}(1+z)$, where $M_0=3\times 10^{10} M_{\odot}$ serves as a reference value. However, due to the absence of high-redshift measurements, our understanding replies on theoretical extrapolation based on the values reported by~\cite{Greene:2019vlv}, which are $\alpha = 7.89\pm 0.09$,  $\beta=1.33\pm 0.12$ and $\epsilon = 0.2$.
In Figure~\ref{contourf} we provide a visual representation of the initial mass, denoted as  $M_{\rm BH,i}$, required to account for the current JWST observations of the high redshift massive galaxies, particularly within the context of the PBH accretion scenario using data obtained from the JWST. And we have assumed that the MBHs at the centers of these high-redshift galaxies observed by JWST all originate from PBHs.

\section{JWST Observations and PBH Mass Distribution}
% \subsection{Observation Data} 
The James Webb Space Telescope (JWST), launched on December 25, 2021, is currently one of the most powerful space telescopes, equipped with four instruments: Mid-Infrared Instrument (MIRI), Near-Infrared Spectrograph (NIRSpec), Near-Infrared Camera (NIRCam) and Fine Guidance Sensor/Near Infrared Imager and Slitless Spectrograph (FGS-NIRISS). During its Early Release Observations (EROs), the JWST observed the massive strong lensing cluster SMACSJ0723. Spectra obtained from this observation revealed both Ly$\alpha$ and Balmer breaks in the galaxies, offering valuable insights into their masses and redshifts through Spectral Energy Distribution (SED) fitting. Furthermore, the Cosmic Evolution Early Release Science (CEERS) program employed NIRCam to capture multi-band images in a deep field, covering an area of approximately $40$ $\rm arcmin^{2}$. These data have been widely utilized in numerous studies to identify candidates for massive high-redshift galaxies using the SED fitting method~\cite{2022arXiv220712446L, 2023MNRAS.519.1201A, 2023ApJ...942L...9Y}. 

However, it is essential to acknowledge that uncertainties persist in the photometric redshift estimation. Therefore, in this analysis, we focus on high-redshift JWST galaxy candidates with spectroscopically determined redshifts, which are  more reliable~\cite{2023Natur.622..707A,2023arXiv230300306B,  2023A&A...677A..88B, 2023NatAs...7..622C, Fujimoto:2023orx, 2023NatAs.tmp..194H, 2023arXiv230405385J}. The galaxies included in Table~\ref{data_table} represent the most prominent candidates identified to date.

% \begin{center}
\begin{table*}
      {\small
      \begin{tabular}{lccccc}
            \hline
            \hline
            \noalign{\smallskip}
             Source ID & $z_{\rm spec}$  & ${\rm log}(M_\star/M_\odot)$ & Reference\\
            \noalign{\smallskip}
            \hline   
            CEERS-61419 & $8.998^{+0.001}_{-0.001}$ & $7.78^{+0.30}_{-0.30}$ & Fu23 \\
            CEERS-61381 & $8.881^{+0.001}_{-0.001}$ & $7.30^{+0.30}_{-0.30}$ & Fu23 \\            
            CEERS-7078 & $8.876^{+0.002}_{-0.002}$ & $8.08^{+0.24}_{-0.30}$ & Fu23 \\            
            CEERS-4702 & $8.807^{+0.003}_{-0.003}$ & $8.08^{+0.22}_{-0.30}$ & Fu23 \\            
            CEERS-4774 & $8.005^{+0.001}_{-0.001}$ & $8.30^{+0.27}_{-0.22}$ & Fu23 \\
            CEERS-4777 & $7.993^{+0.001}_{-0.001}$ & $8.94^{+0.24}_{-0.31}$ & Fu23 \\            
            CEERS-23084 & $7.769^{+0.003}_{-0.003}$ & $9.49^{+0.22}_{-0.24}$ & Fu23 \\
            
            CEERS-43725 & $8.715^{+0.001}_{-0.001}$ & $9.05^{+0.03}_{-0.02}$ & He23 \\
            CEERS-81061 & $8.679^{+0.001}_{-0.001}$ & $10.0^{+0.01}_{-0.01}$ & He23 \\
            EGS-11855 & $8.610^{+0.001}_{-0.001}$ & $9.47^{+0.04}_{-0.06}$ & He23 \\
            EGS-34697 & $8.175^{+0.001}_{-0.001}$ & $9.04^{+0.10}_{-0.11}$ & He23 \\            
            CEERS-59920 & $7.820^{+0.001}_{-0.001}$ & $9.07^{+0.01}_{-0.01}$ & He23 \\
            EGS-8901 & $7.776^{+0.001}_{-0.001}$ & $8.85^{+0.07}_{-0.06}$ & He23 \\

            EGS-33634 & $7.752^{+0.001}_{-0.001}$ & $9.84^{+0.44}_{-0.66}$ & J23 \\
            EGS-36986 & $7.546^{+0.001}_{-0.001}$ & $9.77^{+0.51}_{-0.69}$ & J23 \\ 
            CEERS-16943 & $11.416^{+0.005}_{-0.005}$ & $8.6^{+0.3}_{-0.3}$ & AH23 \\
            CEERS-11384 & $11.043^{+0.003}_{-0.003}$ & $8.7^{+0.1}_{-0.1}$ & AH23 \\

            GS-z10-0 & $10.38^{+0.07}_{-0.06}$ & $7.58^{+0.19}_{-0.20}$ &La23 \\
            GS-z11-0 & $11.58^{+0.05}_{-0.05}$ & $8.67^{+0.08}_{-0.13}$ & La23 \\
            GS-z12-0 & $12.63^{+0.24}_{-0.08}$ & $7.64^{+0.66}_{-0.39}$ & La23 \\
            GS-z13-0 & $13.20^{+0.04}_{-0.07}$ & $7.95^{+0.19}_{-0.29}$ & La23 \\
            
            GN-z11 & $10.603^{0.001}_{0.001}$ & $8.73^{+0.06}_{-0.06}$ & Bu23\\
            Gz9p3 & $9.3127^{+0.0002}_{-0.0002}$ & $9.40^{+0.11}_{-0.10}$ & Bo23\\
           \noalign{\smallskip}
           \hline
           \hline
           \noalign{\smallskip} \noalign{\smallskip}
         \end{tabular} 
         }
\caption{(1) Source ID corresponds to galaxies. (2)Spectroscopic redshift values obtained from measurements of emission lines. (3) Mass of galaxies.(4) Literatures reporting these sources. 
Fu23~\cite{Fujimoto:2023orx}, He23~\cite{2023NatAs.tmp..194H}, J23~\cite{2023arXiv230405385J}, AH23~\cite{2023Natur.622..707A},La23~\cite{2023NatAs...7..622C}, Bu23~\cite{2023A&A...677A..88B}, and Bo23~\cite{2023arXiv230300306B}. }
\label{data_table} 
\end{table*}
% \end{center}

% \subsection{Hierarchical Bayesian Inference and Results}
Furthermore, the Bondi accretion affects the mass distribution of PBHs with redshift. The fraction of PBHs with mass in the interval $(M, M+dM)$ at redshift $z$ is what we refer to as the mass function $\psi(M, z)$. Regarding an initial $\psi(M_i, z_i)$ at formation redshift $z_i$, its evolution is governed by~\cite{DeLuca:2020fpg} 
\begin{equation}
\psi(M_f (M_i, z_f), z_f) dM_f = \psi(M_i, z_i)dM_i.
\end{equation}
where $M_f(M_i, z_f)$ is the PBH final mass at redshift $z_f$.
In the literature, there are theoretical realizations for primordial power spectra that could generate a significantly enhanced power at small scales allowing for an increase in the abundance of massive PBHs \citep{Cai:2023ptf}, such as non-attractor inflation \citep{Kinney:2005vj, Martin:2012pe, Garcia-Bellido:2017mdw, Pi:2021dft} and non-perturbative resonance effects \citep{Cai:2018tuh, Cai:2019jah, Zhou:2020kkf, Cai:2021yvq}. 
To distinguish various forms of theoretically predicted PBH mass functions, we consider the following typical PBH mass functions that arise in PBH formation models \citep{Carr:2020gox, Carr:2020xqk},

\begin{equation}\label{distribution}
\begin{aligned}
\psi_M=\left\{\begin{array}{ll}\vspace{0.3cm}
\frac{1}{\sqrt{2\pi} \sigma M} \exp (-\frac{\rm log^2 (M/M_c)}{2\sigma^2})  & \quad \text{Lognormal},\\ \vspace{0.3cm}
\sum_{n=1} A_n \delta (M-M_{cn}) & \quad \text{Multipeak} , \\ \vspace{0.3cm}
\frac{1}{2}\frac{M_c^{1/2}}{M^{3/2}} \Theta(M-M_c) & \quad \text{Powerlaw}, \\ \vspace{0.3cm}
\frac{1}{\sqrt{2\pi} \sigma } \exp (-\frac{\rm (M-M_c)^2}{2\sigma^2})  & \quad \text{Gaussian},\\ \vspace{0.3cm}
\frac{3.2}{M}\left(\frac{M}{M_{c}}\right)^{3.85} \exp^{-\left(\frac{M}{M_{c}}\right)^{2.85}} & \quad \text{Critical}.
\end{array}\right.
\end{aligned}
\end{equation}
where $\Theta(M-M_c)$ is the step function, and the $M_c, M_{cn}$ and $\sigma$ are parameters in these distributions. For the Multipeak model, we use two normalized Gaussian distributions with same width in our analysis, as the evolution of $\delta (M-M_{cn})$. 

\begin{figure*}%[htbp] 
\centering
\includegraphics[width=0.95\textwidth]{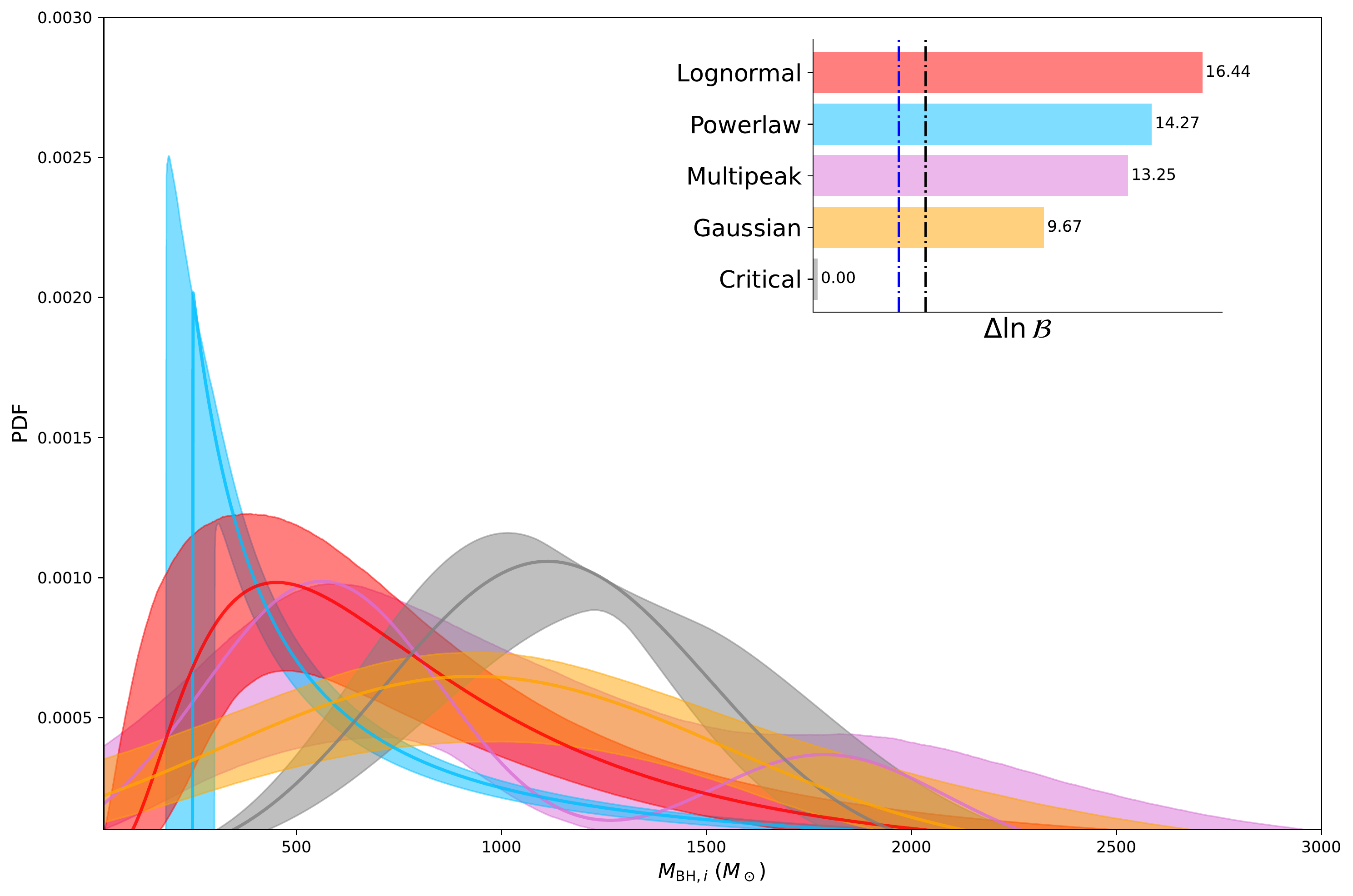}
\caption{The posterior population distribution of the PBH models within a 90\% credible interval. The distribution for Lognormal (red), Multipeak (orchid), Powerlaw (deepskyblue), Gaussian (orange) and Critical (grey) models are shown, with each line plotted using its optimal parameters. The Bayes factors for each model are displayed in the upper right corner, where the vertical dot-dashed lines indicate very strong evidence (blue) and decisive Bayesian evidence (black). Table~\ref{evidence_table} lists the optimal parameters and their 68\% credible intervals.}
\label{model_pdf}
\end{figure*}

\begin{center}
\begin{table*}[htbp]
\setlength{\tabcolsep}{.1em}
\begin{tabular}{lcccccccc}
\hline
 Models & x$_e$ & $M_c$ &  $\sigma$ & $M_{c2}$ & $f_{M_c}$ & $N_{\rm dof}$  & $\Delta \ln\mathcal{B}$ & $ \Delta {\rm AIC}$ \\
\hline
Lognormal & $0.52^{+0.33}_{-0.34}$ & $744.25^{+132.36}_{-111.62}$ &  $0.74^{+0.14}_{-0.11}$ & - & - & 2 & 16.44 & 0.00 \\
\hline
Powerlaw & $0.59^{+0.29}_{-0.37}$ & $243.75^{+36.16}_{-37.03}$ &  - & - & - & 1 & 14.27 & 8.67 \\
\hline
Multipeak & $0.44^{+0.36}_{-0.30}$ & $674.31^{+171.63}_{-152.37}$ &  $462.12^{+170.03}_{-118.10}$ & $1955.31^{+498.69}_{-421.32}$ & $0.77^{+0.13}_{-0.17}$ & 4 & 13.25 & 8.38 \\
\hline
Gaussian & $0.43^{+0.37}_{-0.30}$ & $947.98^{+155.08}_{-149.14}$ &  $694.24^{+135.09}_{-105.36}$ & - & - & 2 & 9.67 & 11.16 \\
\hline
Critical & $0.18^{+0.28}_{-0.13}$ & $1146.58^{+92.51}_{-82.94}$ &  - & - & - & 1 &  0.00 & 31.69\\
\hline
\end{tabular}
\caption{Summary of the results for various initial mass functions of PBH. The first column lists the names of the models, followed by the 68\% credible intervals of their parameters. The last two columns list the Bayes factor $\Delta \ln\mathcal{B}$ and the Akaike information criterion $\Delta \mathrm{AIC}$. Additionally, in~\ref{appendix}, we provide the parameters' priors in Table~\ref{prior} and posterior distributions of individual parameters for five different PBH mass functions in Figure~\ref{five_corner}. }
\label{evidence_table}
\end{table*}
\end{center}

We use the masses of high-redshift galaxies recently observed by JWST to constrain the hyper-parameters $\boldsymbol{\lambda}$ of each initial PBH mass function through hierarchical Bayesian inference~\citep{Thrane:2018qnx}. For a series of $N$ independent observations, the posterior distribution for $\boldsymbol{\lambda}$ is given by

\begin{equation}\label{fun1_hyperparameter}
p(\boldsymbol{\lambda} \mid \boldsymbol{d}) = \pi(\boldsymbol{\lambda}) \prod_{i}^{N}  \int \mathcal{L}({d_i} \mid {\theta}_i) p_{\mathrm{pop}}({\theta_i} \mid \boldsymbol{\lambda}) \mathrm{d} {\theta_i},
\end{equation}
where $\mathcal{L}({d_i} \mid {\theta}_i)$ denote the likelihood function of the JWST data given a galaxy's properties $\theta_i$ (the mass and the redshift). 
The distribution of $\theta_i$ as predicted by the PBH population models is denoted by $p_{\mathrm{pop}}({\theta_i} \mid \boldsymbol{\lambda})$, which satisfies $p_{\mathrm{pop}}({\theta_i} \mid \boldsymbol{\lambda}) = p(m)p(z)$, where $p(m)$ is the mass distribution of galaxies calculated from the initial PBH mass functions in Equation~(\ref{distribution}), and the galaxies are assumed to be distributed uniformly in the co-moving frame of the Universe. We assign Uniform priors $\pi(\boldsymbol{\lambda})$ for all of the hyper-parameters in this work. 
To approximate the reported result for the mass and redshift of each galaxy (as shown by a central value plus/minus the uncertainties in Table~\ref{data_table}), we use a two-dimensional skew normal distribution $\mathcal{N}(\theta_i)$, and assume that $\mathcal{L}({d_i} \mid {\theta}_i) \propto \mathcal{N}(\theta_i)$. Then the above equation can be calculated via Monte Carlo integration with sample points drawn from the skew normal distributions.

To quantitatively evaluate various PBH models and the statistical significance of our results, we calculate the Bayes factor between model $\mathcal{M}_1$ and model $\mathcal{M}_2$, namely $\mathcal{B}^{\mathcal{M}_1}_{\mathcal{M}_2} \equiv Z_{\mathcal{M}_1}/Z_{\mathcal{M}_2}$, where $Z_{\mathcal{M}} \equiv \int \mathrm{d} \boldsymbol{\lambda} p(\boldsymbol{\lambda} \mid \boldsymbol{d})$. 
According to Jeffreys' scale criterion, the Bayes factor $\Delta \mathcal{B}^{\mathcal{M}_1}_{\mathcal{M}_2}$ greater than $(10, 10^{1.5}, 10^2)$ (or $\Delta \ln \mathcal{B}^{\mathcal{M}_1}_{\mathcal{M}_2}$ larger than [2.30, 3.45, 4.60]) would suggest strong, very strong, or decisive Bayesian evidence in favor of model $\mathcal{M}_1$ relative to model $\mathcal{M}_2$, given the available data. 
Additionally, we calculate the Akaike information criterion (AIC)~\citep{Akaike1974ANL}, defining as ${\rm AIC} \equiv 2N_{\rm dof}-2 {\rm ln} (\mathcal{L}({d_i} \mid {\theta}_i))$, to compare models with different numbers of parameters. A difference of $\Delta \text{AIC}$ of 2 or more indicates strong evidence against the model with the higher AIC value.

Our summarized results, presented in Table~\ref{evidence_table}, reveal a preference for the Lognormal, Powerlaw, and Multipeak models over the Gaussian models, with a Bayesian evidence $\ln(\mathcal{B})$ exceeding 3.58. Furthermore, the Lognormal model outperforms the Powerlaw and Multipeak models, and the Powerlaw and Multipeak models exhibit comparable quality as indicated by their slight differences in the Bayesian evidence $\ln(\mathcal{B})$ and AIC values. Notably, the Critical model displays the highest AIC value and the lowest $\ln(\mathcal{B})$, thus being strongly disfavored.
Figure~\ref{model_pdf} displays the constraints on the five different PBH initial mass functions and the Bayes factors of each model compared to the Critical model. The colored shaded zones represent the 90\% credible regions for the inferred population distribution. For all models, the majority of the PBH masses lie $\sim 1000 M_\odot$. According to Jeffreys' scale criterion, the logarithmic Bayes factors for the Lognormal model, the Powerlaw model, the Multipeak model, and the Gasssian model are 16.44, 14.27, 13.25, and 9.67, respectively, indicating they are decisively preferred by the data compared to the Critical model. Most notably, the Lognormal mass function stands out from the five models, with the lowest AIC value of 8.67 compared to the second preferred model. Consequently, we conclude that JWST observations provide informative insights into distinguishing the initial PBH mass function, with current data offering relatively stringent constraints on its shape. Moreover, the above analysis also highlights the importance of statistical analysis in making conclusions about the PBHs population and their implications for early cosmology.

%%%%%%%%%%%%%
\section{Summary} 
The accretion of baryonic matter onto PBHs is believed to have played an important role in the early Universe, giving rise to the formation of seed black holes at the centers of early galaxies. This study demonstrates that PBHs with initial masses of $\sim 1000~M_\odot$ exhibit rapid growth through super-Eddington accretion, reaching masses in the range of $\sim 10^{6}-10^{8}M_\odot$ by $z\gtrsim 7$, which could potentially account for the massive high-redshift galaxies observed by JWST. 
Future multi-band precise observations, such as JWST, SKA and WFST~\cite{WFST:2023voz}, during the cosmic re-ionization era, are expected to reveal these rapidly growing PBHs, as their accretion may also contribute to the radiation background, such as radio and X-ray background~\cite{Ziparo:2022fnc}. 
In our efforts to constrain the formation models of PBHs, we have employed a hierarchical Bayesian inference framework, leading to a strong preference for the Lognormal Model with a characteristic mass of $M_{\rm c}\sim 750M_\odot$ over other scenarios.

It is noteworthy that, in this work, we focus on massive black holes evolved from primordial black holes. However, if the seed black holes were instead formed astrophysically, their mass growth through accretion would not be more efficient. Therefore, the lower bound on the mass of the seed black holes required to account for the JWST observations would still apply. In other words, the seeds must originate as ``intermediate'' mass black holes ($\sim 10^3 M_{\odot}$), regardless of their origins. 
Given that the current high-redshift galaxy sample remains relatively limited, future observation data may modify the current model constraints. Nevertheless, our current results have demonstrated the promising prospect of distinguishing between different PBH distribution models with future observations.

\section*{Acknowledgements} 
We are grateful to Lei Feng, Yu-Yang Songsheng, Jian-Min Wang, Qiang Yuan, Lei Zu for helpful discussion. This work is supported in part by the National Key R\&D Program of China (2021YFC2203100), by the NSFC (11921003, 11961131007, 11653002, 12003029, 12261131497), by China Postdoctoral Science Foundation under grant No. 2023TQ0355, by CAS Young Interdisciplinary Innovation Team (JCTD-2022-20), by 111 Project for "Observational and Theoretical Research on Dark Matter and Dark Energy" (B23042), by the Fundamental Research Funds for Central Universities, by the CSC Innovation Talent Funds, by the CAS project for young scientists in basic research (YSBR-006), and by the USTC Research Funds of the Double First-Class Initiative.

% \bibliographystyle{apsrev}
% \bibliography{reference}

\begin{thebibliography}{64}
\expandafter\ifx\csname natexlab\endcsname\relax\def\natexlab#1{#1}\fi
\expandafter\ifx\csname bibnamefont\endcsname\relax
  \def\bibnamefont#1{#1}\fi
\expandafter\ifx\csname bibfnamefont\endcsname\relax
  \def\bibfnamefont#1{#1}\fi
\expandafter\ifx\csname citenamefont\endcsname\relax
  \def\citenamefont#1{#1}\fi
\expandafter\ifx\csname url\endcsname\relax
  \def\url#1{\texttt{#1}}\fi
\expandafter\ifx\csname urlprefix\endcsname\relax\def\urlprefix{URL }\fi
\providecommand{\bibinfo}[2]{#2}
\providecommand{\eprint}[2][]{\url{#2}}

\bibitem[{\citenamefont{Barkana and Loeb}(2001)}]{Barkana:2000fd}
\bibinfo{author}{\bibfnamefont{R.}~\bibnamefont{Barkana}} \bibnamefont{and}
  \bibinfo{author}{\bibfnamefont{A.}~\bibnamefont{Loeb}},
  \bibinfo{journal}{Phys. Rept.} \textbf{\bibinfo{volume}{349}},
  \bibinfo{pages}{125} (\bibinfo{year}{2001}), \eprint{astro-ph/0010468}.


\bibitem[{\citenamefont{{Bromm} and {Yoshida}}(2011)}]{2011ARA&A..49..373B}
\bibinfo{author}{\bibfnamefont{V.}~\bibnamefont{{Bromm}}} \bibnamefont{and}
  \bibinfo{author}{\bibfnamefont{N.}~\bibnamefont{{Yoshida}}},
  \bibinfo{journal}{\araa} \textbf{\bibinfo{volume}{49}}, \bibinfo{pages}{373}
  (\bibinfo{year}{2011}), \eprint{1102.4638}.

\bibitem[{\citenamefont{Volonteri et~al.}(2021)\citenamefont{Volonteri,
  Habouzit, and Colpi}}]{Volonteri:2021sfo}
\bibinfo{author}{\bibfnamefont{M.}~\bibnamefont{Volonteri}},
  \bibinfo{author}{\bibfnamefont{M.}~\bibnamefont{Habouzit}}, \bibnamefont{and}
  \bibinfo{author}{\bibfnamefont{M.}~\bibnamefont{Colpi}},
  \bibinfo{journal}{Nature Rev. Phys.} \textbf{\bibinfo{volume}{3}},
  \bibinfo{pages}{732} (\bibinfo{year}{2021}), \eprint{2110.10175}.

\bibitem[{\citenamefont{Clesse and Garc\'\i{}a-Bellido}(2015)}]{Clesse:2015wea}
\bibinfo{author}{\bibfnamefont{S.}~\bibnamefont{Clesse}} \bibnamefont{and}
  \bibinfo{author}{\bibfnamefont{J.}~\bibnamefont{Garc\'\i{}a-Bellido}},
  \bibinfo{journal}{Phys. Rev. D} \textbf{\bibinfo{volume}{92}},
  \bibinfo{pages}{023524} (\bibinfo{year}{2015}), \eprint{1501.07565}.

\bibitem[{\citenamefont{Escriv\`a et~al.}(2022)\citenamefont{Escriv\`a, Kuhnel,
  and Tada}}]{Escriva:2022duf}
\bibinfo{author}{\bibfnamefont{A.}~\bibnamefont{Escriv\`a}},
  \bibinfo{author}{\bibfnamefont{F.}~\bibnamefont{Kuhnel}}, \bibnamefont{and}
  \bibinfo{author}{\bibfnamefont{Y.}~\bibnamefont{Tada}}
  (\bibinfo{year}{2022}), \eprint{2211.05767}.

\bibitem[{\citenamefont{{Zel'dovich} and {Novikov}}(1966)}]{1966AZh}
\bibinfo{author}{\bibfnamefont{Y.~B.} \bibnamefont{{Zel'dovich}}}
  \bibnamefont{and} \bibinfo{author}{\bibfnamefont{I.~D.}
  \bibnamefont{{Novikov}}}, \bibinfo{journal}{Astron. Z.}
  \textbf{\bibinfo{volume}{43}}, \bibinfo{pages}{758} (\bibinfo{year}{1966}).

\bibitem[{\citenamefont{{Hawking}}(1971)}]{1971MNRAS}
\bibinfo{author}{\bibfnamefont{S.}~\bibnamefont{{Hawking}}},
  \bibinfo{journal}{\mnras} \textbf{\bibinfo{volume}{152}}, \bibinfo{pages}{75}
  (\bibinfo{year}{1971}).

\bibitem[{\citenamefont{Bertone and Hooper}(2018)}]{Bertone:2016nfn}
\bibinfo{author}{\bibfnamefont{G.}~\bibnamefont{Bertone}} \bibnamefont{and}
  \bibinfo{author}{\bibfnamefont{D.}~\bibnamefont{Hooper}},
  \bibinfo{journal}{Rev. Mod. Phys.} \textbf{\bibinfo{volume}{90}},
  \bibinfo{pages}{045002} (\bibinfo{year}{2018}), \eprint{1605.04909}.

\bibitem[{\citenamefont{Schumann}(2019)}]{Schumann:2019eaa}
\bibinfo{author}{\bibfnamefont{M.}~\bibnamefont{Schumann}},
  \bibinfo{journal}{J. Phys. G} \textbf{\bibinfo{volume}{46}},
  \bibinfo{pages}{103003} (\bibinfo{year}{2019}), \eprint{1903.03026}.

\bibitem[{\citenamefont{{Fan} et~al.}(2022)\citenamefont{{Fan}, {Tang}, {Tsai},
  and {Wu}}}]{2022PhRvL.129i1802F}
\bibinfo{author}{\bibfnamefont{Y.-Z.} \bibnamefont{{Fan}}},
  \bibinfo{author}{\bibfnamefont{T.-P.} \bibnamefont{{Tang}}},
  \bibinfo{author}{\bibfnamefont{Y.-L.~S.} \bibnamefont{{Tsai}}},
  \bibnamefont{and} \bibinfo{author}{\bibfnamefont{L.}~\bibnamefont{{Wu}}},
  \bibinfo{journal}{\prl} \textbf{\bibinfo{volume}{129}}, \bibinfo{eid}{091802}
  (\bibinfo{year}{2022}), \eprint{2204.03693}.

\bibitem[{\citenamefont{{Chapline}}(1975)}]{1975Natur.253..251C}
\bibinfo{author}{\bibfnamefont{G.~F.} \bibnamefont{{Chapline}}},
  \bibinfo{journal}{\nat} \textbf{\bibinfo{volume}{253}}, \bibinfo{pages}{251}
  (\bibinfo{year}{1975}).

\bibitem[{\citenamefont{Carr et~al.}(2016)\citenamefont{Carr, Kuhnel, and
  Sandstad}}]{Carr:2016drx}
\bibinfo{author}{\bibfnamefont{B.}~\bibnamefont{Carr}},
  \bibinfo{author}{\bibfnamefont{F.}~\bibnamefont{Kuhnel}}, \bibnamefont{and}
  \bibinfo{author}{\bibfnamefont{M.}~\bibnamefont{Sandstad}},
  \bibinfo{journal}{Phys. Rev. D} \textbf{\bibinfo{volume}{94}},
  \bibinfo{pages}{083504} (\bibinfo{year}{2016}), \eprint{1607.06077}.

\bibitem[{\citenamefont{Carr et~al.}(2021)\citenamefont{Carr, Kohri, Sendouda,
  and Yokoyama}}]{Carr:2020gox}
\bibinfo{author}{\bibfnamefont{B.}~\bibnamefont{Carr}},
  \bibinfo{author}{\bibfnamefont{K.}~\bibnamefont{Kohri}},
  \bibinfo{author}{\bibfnamefont{Y.}~\bibnamefont{Sendouda}}, \bibnamefont{and}
  \bibinfo{author}{\bibfnamefont{J.}~\bibnamefont{Yokoyama}},
  \bibinfo{journal}{Rept. Prog. Phys.} \textbf{\bibinfo{volume}{84}},
  \bibinfo{pages}{116902} (\bibinfo{year}{2021}), \eprint{2002.12778}.

\bibitem[{\citenamefont{Inayoshi et~al.}(2020)\citenamefont{Inayoshi, Visbal,
  and Haiman}}]{Inayoshi:2019fun}
\bibinfo{author}{\bibfnamefont{K.}~\bibnamefont{Inayoshi}},
  \bibinfo{author}{\bibfnamefont{E.}~\bibnamefont{Visbal}}, \bibnamefont{and}
  \bibinfo{author}{\bibfnamefont{Z.}~\bibnamefont{Haiman}},
  \bibinfo{journal}{Ann. Rev. Astron. Astrophys.}
  \textbf{\bibinfo{volume}{58}}, \bibinfo{pages}{27} (\bibinfo{year}{2020}),
  \eprint{1911.05791}.

\bibitem[{\citenamefont{Gardner et~al.}(2006)}]{Gardner:2006ky}
\bibinfo{author}{\bibfnamefont{J.~P.} \bibnamefont{Gardner}}
  \bibnamefont{et~al.}, \bibinfo{journal}{Space Sci. Rev.}
  \textbf{\bibinfo{volume}{123}}, \bibinfo{pages}{485} (\bibinfo{year}{2006}),
  \eprint{astro-ph/0606175}.

\bibitem[{\citenamefont{{Lu} et~al.}(2019)\citenamefont{{Lu}, {Huang}, {Zhang},
  {Wang}, {Du}, {Hu}, {Xiao}, {Li}, {Bai}, {Bian}
  et~al.}}]{2019ApJ...877...23L}
\bibinfo{author}{\bibfnamefont{K.-X.} \bibnamefont{{Lu}}},
  \bibinfo{author}{\bibfnamefont{Y.-K.} \bibnamefont{{Huang}}},
  \bibinfo{author}{\bibfnamefont{Z.-X.} \bibnamefont{{Zhang}}},
  \bibinfo{author}{\bibfnamefont{K.}~\bibnamefont{{Wang}}},
  \bibinfo{author}{\bibfnamefont{P.}~\bibnamefont{{Du}}},
  \bibinfo{author}{\bibfnamefont{C.}~\bibnamefont{{Hu}}},
  \bibinfo{author}{\bibfnamefont{M.}~\bibnamefont{{Xiao}}},
  \bibinfo{author}{\bibfnamefont{Y.-R.} \bibnamefont{{Li}}},
  \bibinfo{author}{\bibfnamefont{J.-M.} \bibnamefont{{Bai}}},
  \bibinfo{author}{\bibfnamefont{W.-H.} \bibnamefont{{Bian}}},
  \bibnamefont{et~al.}, \bibinfo{journal}{\apj} \textbf{\bibinfo{volume}{877}},
  \bibinfo{eid}{23} (\bibinfo{year}{2019}), \eprint{1904.03393}.

\bibitem[{\citenamefont{Massonneau et~al.}(2023)\citenamefont{Massonneau,
  Volonteri, Dubois, and Beckmann}}]{Massonneau:2022sye}
\bibinfo{author}{\bibfnamefont{W.}~\bibnamefont{Massonneau}},
  \bibinfo{author}{\bibfnamefont{M.}~\bibnamefont{Volonteri}},
  \bibinfo{author}{\bibfnamefont{Y.}~\bibnamefont{Dubois}}, \bibnamefont{and}
  \bibinfo{author}{\bibfnamefont{R.~S.} \bibnamefont{Beckmann}},
  \bibinfo{journal}{Astron. Astrophys.} \textbf{\bibinfo{volume}{670}},
  \bibinfo{pages}{A180} (\bibinfo{year}{2023}), \eprint{2201.08766}.

\bibitem[{\citenamefont{{Labbe} et~al.}(2022)\citenamefont{{Labbe}, {van
  Dokkum}, {Nelson}, {Bezanson}, {Suess}, {Leja}, {Brammer}, {Whitaker},
  {Mathews}, {Stefanon} et~al.}}]{2022arXiv220712446L}
\bibinfo{author}{\bibfnamefont{I.}~\bibnamefont{{Labbe}}},
  \bibinfo{author}{\bibfnamefont{P.}~\bibnamefont{{van Dokkum}}},
  \bibinfo{author}{\bibfnamefont{E.}~\bibnamefont{{Nelson}}},
  \bibinfo{author}{\bibfnamefont{R.}~\bibnamefont{{Bezanson}}},
  \bibinfo{author}{\bibfnamefont{K.}~\bibnamefont{{Suess}}},
  \bibinfo{author}{\bibfnamefont{J.}~\bibnamefont{{Leja}}},
  \bibinfo{author}{\bibfnamefont{G.}~\bibnamefont{{Brammer}}},
  \bibinfo{author}{\bibfnamefont{K.}~\bibnamefont{{Whitaker}}},
  \bibinfo{author}{\bibfnamefont{E.}~\bibnamefont{{Mathews}}},
  \bibinfo{author}{\bibfnamefont{M.}~\bibnamefont{{Stefanon}}},
  \bibnamefont{et~al.}, \bibinfo{journal}{Nature,}
  \bibinfo{eid}{arXiv:2207.12446} (\bibinfo{year}{2022}), \eprint{2207.12446}.

\bibitem[{\citenamefont{{Atek} et~al.}(2023)\citenamefont{{Atek}, {Shuntov},
  {Furtak}, {Richard}, {Kneib}, {Mahler}, {Zitrin}, {McCracken}, {Charlot},
  {Chevallard} et~al.}}]{2023MNRAS.519.1201A}
\bibinfo{author}{\bibfnamefont{H.}~\bibnamefont{{Atek}}},
  \bibinfo{author}{\bibfnamefont{M.}~\bibnamefont{{Shuntov}}},
  \bibinfo{author}{\bibfnamefont{L.~J.} \bibnamefont{{Furtak}}},
  \bibinfo{author}{\bibfnamefont{J.}~\bibnamefont{{Richard}}},
  \bibinfo{author}{\bibfnamefont{J.-P.} \bibnamefont{{Kneib}}},
  \bibinfo{author}{\bibfnamefont{G.}~\bibnamefont{{Mahler}}},
  \bibinfo{author}{\bibfnamefont{A.}~\bibnamefont{{Zitrin}}},
  \bibinfo{author}{\bibfnamefont{H.~J.} \bibnamefont{{McCracken}}},
  \bibinfo{author}{\bibfnamefont{S.}~\bibnamefont{{Charlot}}},
  \bibinfo{author}{\bibfnamefont{J.}~\bibnamefont{{Chevallard}}},
  \bibnamefont{et~al.}, \bibinfo{journal}{\mnras}
  \textbf{\bibinfo{volume}{519}}, \bibinfo{pages}{1201} (\bibinfo{year}{2023}),
  \eprint{2207.12338}.

\bibitem[{\citenamefont{{Yan} et~al.}(2023)\citenamefont{{Yan}, {Ma}, {Ling},
  {Cheng}, and {Huang}}}]{2023ApJ...942L...9Y}
\bibinfo{author}{\bibfnamefont{H.}~\bibnamefont{{Yan}}},
  \bibinfo{author}{\bibfnamefont{Z.}~\bibnamefont{{Ma}}},
  \bibinfo{author}{\bibfnamefont{C.}~\bibnamefont{{Ling}}},
  \bibinfo{author}{\bibfnamefont{C.}~\bibnamefont{{Cheng}}}, \bibnamefont{and}
  \bibinfo{author}{\bibfnamefont{J.-S.} \bibnamefont{{Huang}}},
  \bibinfo{journal}{\apjl} \textbf{\bibinfo{volume}{942}}, \bibinfo{eid}{L9}
  (\bibinfo{year}{2023}), \eprint{2207.11558}.

\bibitem[{\citenamefont{{Lovell} et~al.}(2023)\citenamefont{{Lovell},
  {Harrison}, {Harikane}, {Tacchella}, and {Wilkins}}}]{2023MNRAS.518.2511L}
\bibinfo{author}{\bibfnamefont{C.~C.} \bibnamefont{{Lovell}}},
  \bibinfo{author}{\bibfnamefont{I.}~\bibnamefont{{Harrison}}},
  \bibinfo{author}{\bibfnamefont{Y.}~\bibnamefont{{Harikane}}},
  \bibinfo{author}{\bibfnamefont{S.}~\bibnamefont{{Tacchella}}},
  \bibnamefont{and} \bibinfo{author}{\bibfnamefont{S.~M.}
  \bibnamefont{{Wilkins}}}, \bibinfo{journal}{\mnras}
  \textbf{\bibinfo{volume}{518}}, \bibinfo{pages}{2511} (\bibinfo{year}{2023}),
  \eprint{2208.10479}.

\bibitem[{\citenamefont{Liu and Bromm}(2022)}]{Liu:2022bvr}
\bibinfo{author}{\bibfnamefont{B.}~\bibnamefont{Liu}} \bibnamefont{and}
  \bibinfo{author}{\bibfnamefont{V.}~\bibnamefont{Bromm}},
  \bibinfo{journal}{Astrophys. J. Lett.} \textbf{\bibinfo{volume}{937}},
  \bibinfo{pages}{L30} (\bibinfo{year}{2022}), \eprint{2208.13178}.

\bibitem[{\citenamefont{{Biagetti} et~al.}(2023)\citenamefont{{Biagetti},
  {Franciolini}, and {Riotto}}}]{2023ApJ...944..113B}
\bibinfo{author}{\bibfnamefont{M.}~\bibnamefont{{Biagetti}}},
  \bibinfo{author}{\bibfnamefont{G.}~\bibnamefont{{Franciolini}}},
  \bibnamefont{and} \bibinfo{author}{\bibfnamefont{A.}~\bibnamefont{{Riotto}}},
  \bibinfo{journal}{\apj} \textbf{\bibinfo{volume}{944}}, \bibinfo{eid}{113}
  (\bibinfo{year}{2023}), \eprint{2210.04812}.

\bibitem[{\citenamefont{{Trinca} et~al.}(2023)\citenamefont{{Trinca},
  {Schneider}, {Maiolino}, {Valiante}, {Graziani}, and
  {Volonteri}}}]{2023MNRAS.519.4753T}
\bibinfo{author}{\bibfnamefont{A.}~\bibnamefont{{Trinca}}},
  \bibinfo{author}{\bibfnamefont{R.}~\bibnamefont{{Schneider}}},
  \bibinfo{author}{\bibfnamefont{R.}~\bibnamefont{{Maiolino}}},
  \bibinfo{author}{\bibfnamefont{R.}~\bibnamefont{{Valiante}}},
  \bibinfo{author}{\bibfnamefont{L.}~\bibnamefont{{Graziani}}},
  \bibnamefont{and}
  \bibinfo{author}{\bibfnamefont{M.}~\bibnamefont{{Volonteri}}},
  \bibinfo{journal}{\mnras} \textbf{\bibinfo{volume}{519}},
  \bibinfo{pages}{4753} (\bibinfo{year}{2023}), \eprint{2211.01389}.

\bibitem[{\citenamefont{{H{\"u}tsi} et~al.}(2023)\citenamefont{{H{\"u}tsi},
  {Raidal}, {Urrutia}, {Vaskonen}, and {Veerm{\"a}e}}}]{2023PhRvD.107d3502H}
\bibinfo{author}{\bibfnamefont{G.}~\bibnamefont{{H{\"u}tsi}}},
  \bibinfo{author}{\bibfnamefont{M.}~\bibnamefont{{Raidal}}},
  \bibinfo{author}{\bibfnamefont{J.}~\bibnamefont{{Urrutia}}},
  \bibinfo{author}{\bibfnamefont{V.}~\bibnamefont{{Vaskonen}}},
  \bibnamefont{and}
  \bibinfo{author}{\bibfnamefont{H.}~\bibnamefont{{Veerm{\"a}e}}},
  \bibinfo{journal}{\prd} \textbf{\bibinfo{volume}{107}}, \bibinfo{eid}{043502}
  (\bibinfo{year}{2023}), \eprint{2211.02651}.

\bibitem[{\citenamefont{{Cai} et~al.}(2023)\citenamefont{{Cai}, {Tang}, {Mo},
  {Yan}, {Chen}, {Ma}, {Wang}, {Luo}, {Easson}, and
  {Marciano}}}]{2023arXiv230109403C}
\bibinfo{author}{\bibfnamefont{Y.-F.} \bibnamefont{{Cai}}},
  \bibinfo{author}{\bibfnamefont{C.}~\bibnamefont{{Tang}}},
  \bibinfo{author}{\bibfnamefont{G.}~\bibnamefont{{Mo}}},
  \bibinfo{author}{\bibfnamefont{S.}~\bibnamefont{{Yan}}},
  \bibinfo{author}{\bibfnamefont{C.}~\bibnamefont{{Chen}}},
  \bibinfo{author}{\bibfnamefont{X.}~\bibnamefont{{Ma}}},
  \bibinfo{author}{\bibfnamefont{B.}~\bibnamefont{{Wang}}},
  \bibinfo{author}{\bibfnamefont{W.}~\bibnamefont{{Luo}}},
  \bibinfo{author}{\bibfnamefont{D.}~\bibnamefont{{Easson}}}, \bibnamefont{and}
  \bibinfo{author}{\bibfnamefont{A.}~\bibnamefont{{Marciano}}},
  \bibinfo{journal}{arXiv e-prints} \bibinfo{eid}{arXiv:2301.09403}
  (\bibinfo{year}{2023}), \eprint{2301.09403}.

\bibitem[{\citenamefont{Wang et~al.}(2023{\natexlab{a}})\citenamefont{Wang,
  Lei, Jiao, Feng, and Fan}}]{Wang:2023len}
\bibinfo{author}{\bibfnamefont{Z.}~\bibnamefont{Wang}},
  \bibinfo{author}{\bibfnamefont{L.}~\bibnamefont{Lei}},
  \bibinfo{author}{\bibfnamefont{H.}~\bibnamefont{Jiao}},
  \bibinfo{author}{\bibfnamefont{L.}~\bibnamefont{Feng}}, \bibnamefont{and}
  \bibinfo{author}{\bibfnamefont{Y.-Z.} \bibnamefont{Fan}},
  \bibinfo{journal}{Sci. China Phys. Mech. Astron.}
  \textbf{\bibinfo{volume}{66}}, \bibinfo{pages}{120403}
  (\bibinfo{year}{2023}{\natexlab{a}}), \eprint{2306.17150}.

\bibitem[{\citenamefont{Wang et~al.}(2013)\citenamefont{Wang, Du, Valls-Gabaud,
  Hu, and Netzer}}]{Wang:2013ha}
\bibinfo{author}{\bibfnamefont{J.~M.} \bibnamefont{Wang}},
  \bibinfo{author}{\bibfnamefont{P.}~\bibnamefont{Du}},
  \bibinfo{author}{\bibfnamefont{D.}~\bibnamefont{Valls-Gabaud}},
  \bibinfo{author}{\bibfnamefont{C.}~\bibnamefont{Hu}}, \bibnamefont{and}
  \bibinfo{author}{\bibfnamefont{H.}~\bibnamefont{Netzer}},
  \bibinfo{journal}{Phys. Rev. Lett.} \textbf{\bibinfo{volume}{110}},
  \bibinfo{pages}{081301} (\bibinfo{year}{2013}), \eprint{1301.4225}.

\bibitem[{\citenamefont{De~Luca
  et~al.}(2020{\natexlab{a}})\citenamefont{De~Luca, Franciolini, Pani, and
  Riotto}}]{DeLuca:2020fpg}
\bibinfo{author}{\bibfnamefont{V.}~\bibnamefont{De~Luca}},
  \bibinfo{author}{\bibfnamefont{G.}~\bibnamefont{Franciolini}},
  \bibinfo{author}{\bibfnamefont{P.}~\bibnamefont{Pani}}, \bibnamefont{and}
  \bibinfo{author}{\bibfnamefont{A.}~\bibnamefont{Riotto}},
  \bibinfo{journal}{Phys. Rev. D} \textbf{\bibinfo{volume}{102}},
  \bibinfo{pages}{043505} (\bibinfo{year}{2020}{\natexlab{a}}),
  \eprint{2003.12589}.

\bibitem[{\citenamefont{Serpico et~al.}(2020)\citenamefont{Serpico, Poulin,
  Inman, and Kohri}}]{Serpico:2020ehh}
\bibinfo{author}{\bibfnamefont{P.~D.} \bibnamefont{Serpico}},
  \bibinfo{author}{\bibfnamefont{V.}~\bibnamefont{Poulin}},
  \bibinfo{author}{\bibfnamefont{D.}~\bibnamefont{Inman}}, \bibnamefont{and}
  \bibinfo{author}{\bibfnamefont{K.}~\bibnamefont{Kohri}},
  \bibinfo{journal}{Phys. Rev. Res.} \textbf{\bibinfo{volume}{2}},
  \bibinfo{pages}{023204} (\bibinfo{year}{2020}), \eprint{2002.10771}.

\bibitem[{\citenamefont{Hasinger}(2020)}]{Hasinger:2020ptw}
\bibinfo{author}{\bibfnamefont{G.}~\bibnamefont{Hasinger}},
  \bibinfo{journal}{JCAP} \textbf{\bibinfo{volume}{07}}, \bibinfo{pages}{022}
  (\bibinfo{year}{2020}), \eprint{2003.05150}.

\bibitem[{\citenamefont{Mack et~al.}(2007)\citenamefont{Mack, Ostriker, and
  Ricotti}}]{Mack:2006gz}
\bibinfo{author}{\bibfnamefont{K.~J.} \bibnamefont{Mack}},
  \bibinfo{author}{\bibfnamefont{J.~P.} \bibnamefont{Ostriker}},
  \bibnamefont{and} \bibinfo{author}{\bibfnamefont{M.}~\bibnamefont{Ricotti}},
  \bibinfo{journal}{Astrophys. J.} \textbf{\bibinfo{volume}{665}},
  \bibinfo{pages}{1277} (\bibinfo{year}{2007}), \eprint{astro-ph/0608642}.

\bibitem[{\citenamefont{Ricotti}(2007)}]{Ricotti:2007jk}
\bibinfo{author}{\bibfnamefont{M.}~\bibnamefont{Ricotti}},
  \bibinfo{journal}{Astrophys. J.} \textbf{\bibinfo{volume}{662}},
  \bibinfo{pages}{53} (\bibinfo{year}{2007}), \eprint{0706.0864}.

\bibitem[{\citenamefont{Ricotti et~al.}(2008)\citenamefont{Ricotti, Ostriker,
  and Mack}}]{Ricotti:2007au}
\bibinfo{author}{\bibfnamefont{M.}~\bibnamefont{Ricotti}},
  \bibinfo{author}{\bibfnamefont{J.~P.} \bibnamefont{Ostriker}},
  \bibnamefont{and} \bibinfo{author}{\bibfnamefont{K.~J.} \bibnamefont{Mack}},
  \bibinfo{journal}{Astrophys. J.} \textbf{\bibinfo{volume}{680}},
  \bibinfo{pages}{829} (\bibinfo{year}{2008}), \eprint{0709.0524}.

\bibitem[{\citenamefont{{Bertschinger}}(1985)}]{1985ApJS...58...39B}
\bibinfo{author}{\bibfnamefont{E.}~\bibnamefont{{Bertschinger}}},
  \bibinfo{journal}{\apjs} \textbf{\bibinfo{volume}{58}}, \bibinfo{pages}{39}
  (\bibinfo{year}{1985}).

\bibitem[{\citenamefont{Ali-Ha\"\i{}moud and
  Kamionkowski}(2017)}]{Ali-Haimoud:2016mbv}
\bibinfo{author}{\bibfnamefont{Y.}~\bibnamefont{Ali-Ha\"\i{}moud}}
  \bibnamefont{and}
  \bibinfo{author}{\bibfnamefont{M.}~\bibnamefont{Kamionkowski}},
  \bibinfo{journal}{Phys. Rev. D} \textbf{\bibinfo{volume}{95}},
  \bibinfo{pages}{043534} (\bibinfo{year}{2017}), \eprint{1612.05644}.

\bibitem[{\citenamefont{De~Luca
  et~al.}(2020{\natexlab{b}})\citenamefont{De~Luca, Franciolini, Pani, and
  Riotto}}]{DeLuca:2020bjf}
\bibinfo{author}{\bibfnamefont{V.}~\bibnamefont{De~Luca}},
  \bibinfo{author}{\bibfnamefont{G.}~\bibnamefont{Franciolini}},
  \bibinfo{author}{\bibfnamefont{P.}~\bibnamefont{Pani}}, \bibnamefont{and}
  \bibinfo{author}{\bibfnamefont{A.}~\bibnamefont{Riotto}},
  \bibinfo{journal}{JCAP} \textbf{\bibinfo{volume}{04}}, \bibinfo{pages}{052}
  (\bibinfo{year}{2020}{\natexlab{b}}), \eprint{2003.02778}.

\bibitem[{\citenamefont{Bosch-Ramon and Bellomo}(2020)}]{Bosch-Ramon:2020pcz}
\bibinfo{author}{\bibfnamefont{V.}~\bibnamefont{Bosch-Ramon}} \bibnamefont{and}
  \bibinfo{author}{\bibfnamefont{N.}~\bibnamefont{Bellomo}},
  \bibinfo{journal}{Astron. Astrophys.} \textbf{\bibinfo{volume}{638}},
  \bibinfo{pages}{A132} (\bibinfo{year}{2020}), \eprint{2004.11224}.

\bibitem[{\citenamefont{Bosch-Ramon}(2022)}]{Bosch-Ramon:2022eiy}
\bibinfo{author}{\bibfnamefont{V.}~\bibnamefont{Bosch-Ramon}},
  \bibinfo{journal}{Astron. Astrophys.} \textbf{\bibinfo{volume}{660}},
  \bibinfo{pages}{A5} (\bibinfo{year}{2022}), \eprint{2201.09601}.

\bibitem[{\citenamefont{Piga et~al.}(2022)\citenamefont{Piga, Lucca, Bellomo,
  Bosch-Ramon, Matarrese, Raccanelli, and Verde}}]{Piga:2022ysp}
\bibinfo{author}{\bibfnamefont{L.}~\bibnamefont{Piga}},
  \bibinfo{author}{\bibfnamefont{M.}~\bibnamefont{Lucca}},
  \bibinfo{author}{\bibfnamefont{N.}~\bibnamefont{Bellomo}},
  \bibinfo{author}{\bibfnamefont{V.}~\bibnamefont{Bosch-Ramon}},
  \bibinfo{author}{\bibfnamefont{S.}~\bibnamefont{Matarrese}},
  \bibinfo{author}{\bibfnamefont{A.}~\bibnamefont{Raccanelli}},
  \bibnamefont{and} \bibinfo{author}{\bibfnamefont{L.}~\bibnamefont{Verde}},
  \bibinfo{journal}{JCAP} \textbf{\bibinfo{volume}{12}}, \bibinfo{pages}{016}
  (\bibinfo{year}{2022}), \eprint{2210.14934}.

\bibitem[{\citenamefont{{Kohri} et~al.}(2022)\citenamefont{{Kohri},
  {Sekiguchi}, and {Wang}}}]{2022PhRvD.106d3539K}
\bibinfo{author}{\bibfnamefont{K.}~\bibnamefont{{Kohri}}},
  \bibinfo{author}{\bibfnamefont{T.}~\bibnamefont{{Sekiguchi}}},
  \bibnamefont{and} \bibinfo{author}{\bibfnamefont{S.}~\bibnamefont{{Wang}}},
  \bibinfo{journal}{\prd} \textbf{\bibinfo{volume}{106}}, \bibinfo{eid}{043539}
  (\bibinfo{year}{2022}), \eprint{2201.05300}.

\bibitem[{\citenamefont{Greene et~al.}(2020)\citenamefont{Greene, Strader, and
  Ho}}]{Greene:2019vlv}
\bibinfo{author}{\bibfnamefont{J.~E.} \bibnamefont{Greene}},
  \bibinfo{author}{\bibfnamefont{J.}~\bibnamefont{Strader}}, \bibnamefont{and}
  \bibinfo{author}{\bibfnamefont{L.~C.} \bibnamefont{Ho}},
  \bibinfo{journal}{Ann. Rev. Astron. Astrophys.}
  \textbf{\bibinfo{volume}{58}}, \bibinfo{pages}{257} (\bibinfo{year}{2020}),
  \eprint{1911.09678}.

\bibitem[{\citenamefont{{Arrabal Haro} et~al.}(2023)\citenamefont{{Arrabal
  Haro}, {Dickinson}, {Finkelstein}, {Kartaltepe}, {Donnan}, {Burgarella},
  {Carnall}, {Cullen}, {Dunlop}, {Fern{\'a}ndez} et~al.}}]{2023Natur.622..707A}
\bibinfo{author}{\bibfnamefont{P.}~\bibnamefont{{Arrabal Haro}}},
  \bibinfo{author}{\bibfnamefont{M.}~\bibnamefont{{Dickinson}}},
  \bibinfo{author}{\bibfnamefont{S.~L.} \bibnamefont{{Finkelstein}}},
  \bibinfo{author}{\bibfnamefont{J.~S.} \bibnamefont{{Kartaltepe}}},
  \bibinfo{author}{\bibfnamefont{C.~T.} \bibnamefont{{Donnan}}},
  \bibinfo{author}{\bibfnamefont{D.}~\bibnamefont{{Burgarella}}},
  \bibinfo{author}{\bibfnamefont{A.~C.} \bibnamefont{{Carnall}}},
  \bibinfo{author}{\bibfnamefont{F.}~\bibnamefont{{Cullen}}},
  \bibinfo{author}{\bibfnamefont{J.~S.} \bibnamefont{{Dunlop}}},
  \bibinfo{author}{\bibfnamefont{V.}~\bibnamefont{{Fern{\'a}ndez}}},
  \bibnamefont{et~al.}, \bibinfo{journal}{\nat} \textbf{\bibinfo{volume}{622}},
  \bibinfo{pages}{707} (\bibinfo{year}{2023}), \eprint{2303.15431}.

\bibitem[{\citenamefont{{Boyett} et~al.}(2023)\citenamefont{{Boyett}, {Trenti},
  {Leethochawalit}, {Calabr{\'o}}, {Metha}, {Roberts-Borsani}, {Dalmasso},
  {Yang}, {Santini}, {Treu} et~al.}}]{2023arXiv230300306B}
\bibinfo{author}{\bibfnamefont{K.}~\bibnamefont{{Boyett}}},
  \bibinfo{author}{\bibfnamefont{M.}~\bibnamefont{{Trenti}}},
  \bibinfo{author}{\bibfnamefont{N.}~\bibnamefont{{Leethochawalit}}},
  \bibinfo{author}{\bibfnamefont{A.}~\bibnamefont{{Calabr{\'o}}}},
  \bibinfo{author}{\bibfnamefont{B.}~\bibnamefont{{Metha}}},
  \bibinfo{author}{\bibfnamefont{G.}~\bibnamefont{{Roberts-Borsani}}},
  \bibinfo{author}{\bibfnamefont{N.}~\bibnamefont{{Dalmasso}}},
  \bibinfo{author}{\bibfnamefont{L.}~\bibnamefont{{Yang}}},
  \bibinfo{author}{\bibfnamefont{P.}~\bibnamefont{{Santini}}},
  \bibinfo{author}{\bibfnamefont{T.}~\bibnamefont{{Treu}}},
  \bibnamefont{et~al.}, \bibinfo{journal}{arXiv e-prints}
  \bibinfo{eid}{arXiv:2303.00306} (\bibinfo{year}{2023}), \eprint{2303.00306}.

\bibitem[{\citenamefont{{Bunker} et~al.}(2023)\citenamefont{{Bunker}, {Saxena},
  {Cameron}, {Willott}, {Curtis-Lake}, {Jakobsen}, {Carniani}, {Smit},
  {Maiolino}, {Witstok} et~al.}}]{2023A&A...677A..88B}
\bibinfo{author}{\bibfnamefont{A.~J.} \bibnamefont{{Bunker}}},
  \bibinfo{author}{\bibfnamefont{A.}~\bibnamefont{{Saxena}}},
  \bibinfo{author}{\bibfnamefont{A.~J.} \bibnamefont{{Cameron}}},
  \bibinfo{author}{\bibfnamefont{C.~J.} \bibnamefont{{Willott}}},
  \bibinfo{author}{\bibfnamefont{E.}~\bibnamefont{{Curtis-Lake}}},
  \bibinfo{author}{\bibfnamefont{P.}~\bibnamefont{{Jakobsen}}},
  \bibinfo{author}{\bibfnamefont{S.}~\bibnamefont{{Carniani}}},
  \bibinfo{author}{\bibfnamefont{R.}~\bibnamefont{{Smit}}},
  \bibinfo{author}{\bibfnamefont{R.}~\bibnamefont{{Maiolino}}},
  \bibinfo{author}{\bibfnamefont{J.}~\bibnamefont{{Witstok}}},
  \bibnamefont{et~al.}, \bibinfo{journal}{\aap} \textbf{\bibinfo{volume}{677}},
  \bibinfo{eid}{A88} (\bibinfo{year}{2023}), \eprint{2302.07256}.

\bibitem[{\citenamefont{{Curtis-Lake} et~al.}(2023)\citenamefont{{Curtis-Lake},
  {Carniani}, {Cameron}, {Charlot}, {Jakobsen}, {Maiolino}, {Bunker},
  {Witstok}, {Smit}, {Chevallard} et~al.}}]{2023NatAs...7..622C}
\bibinfo{author}{\bibfnamefont{E.}~\bibnamefont{{Curtis-Lake}}},
  \bibinfo{author}{\bibfnamefont{S.}~\bibnamefont{{Carniani}}},
  \bibinfo{author}{\bibfnamefont{A.}~\bibnamefont{{Cameron}}},
  \bibinfo{author}{\bibfnamefont{S.}~\bibnamefont{{Charlot}}},
  \bibinfo{author}{\bibfnamefont{P.}~\bibnamefont{{Jakobsen}}},
  \bibinfo{author}{\bibfnamefont{R.}~\bibnamefont{{Maiolino}}},
  \bibinfo{author}{\bibfnamefont{A.}~\bibnamefont{{Bunker}}},
  \bibinfo{author}{\bibfnamefont{J.}~\bibnamefont{{Witstok}}},
  \bibinfo{author}{\bibfnamefont{R.}~\bibnamefont{{Smit}}},
  \bibinfo{author}{\bibfnamefont{J.}~\bibnamefont{{Chevallard}}},
  \bibnamefont{et~al.}, \bibinfo{journal}{Nature Astronomy}
  \textbf{\bibinfo{volume}{7}}, \bibinfo{pages}{622} (\bibinfo{year}{2023}),
  \eprint{2212.04568}.

\bibitem[{\citenamefont{Fujimoto et~al.}(2023)}]{Fujimoto:2023orx}
\bibinfo{author}{\bibfnamefont{S.}~\bibnamefont{Fujimoto}}
  \bibnamefont{et~al.}, \bibinfo{journal}{Astrophys. J. Lett.}
  \textbf{\bibinfo{volume}{949}}, \bibinfo{pages}{L25} (\bibinfo{year}{2023}),
  \eprint{2301.09482}.

\bibitem[{\citenamefont{{Heintz} et~al.}(2023)\citenamefont{{Heintz},
  {Brammer}, {Gim{\'e}nez-Arteaga}, {Strait}, {del P. Lagos}, {Vijayan},
  {Matthee}, {Watson}, {Mason}, {Hutter} et~al.}}]{2023NatAs.tmp..194H}
\bibinfo{author}{\bibfnamefont{K.~E.} \bibnamefont{{Heintz}}},
  \bibinfo{author}{\bibfnamefont{G.~B.} \bibnamefont{{Brammer}}},
  \bibinfo{author}{\bibfnamefont{C.}~\bibnamefont{{Gim{\'e}nez-Arteaga}}},
  \bibinfo{author}{\bibfnamefont{V.~B.} \bibnamefont{{Strait}}},
  \bibinfo{author}{\bibfnamefont{C.}~\bibnamefont{{del P. Lagos}}},
  \bibinfo{author}{\bibfnamefont{A.~P.} \bibnamefont{{Vijayan}}},
  \bibinfo{author}{\bibfnamefont{J.}~\bibnamefont{{Matthee}}},
  \bibinfo{author}{\bibfnamefont{D.}~\bibnamefont{{Watson}}},
  \bibinfo{author}{\bibfnamefont{C.~A.} \bibnamefont{{Mason}}},
  \bibinfo{author}{\bibfnamefont{A.}~\bibnamefont{{Hutter}}},
  \bibnamefont{et~al.}, \bibinfo{journal}{Nature Astronomy}
  (\bibinfo{year}{2023}), \eprint{2212.02890}.

\bibitem[{\citenamefont{{Jung} et~al.}(2023)\citenamefont{{Jung},
  {Finkelstein}, {Arrabal Haro}, {Dickinson}, {Ferguson}, {Hutchison},
  {Kartaltepe}, {Larson}, {Simons}, {Papovich} et~al.}}]{2023arXiv230405385J}
\bibinfo{author}{\bibfnamefont{I.}~\bibnamefont{{Jung}}},
  \bibinfo{author}{\bibfnamefont{S.~L.} \bibnamefont{{Finkelstein}}},
  \bibinfo{author}{\bibfnamefont{P.}~\bibnamefont{{Arrabal Haro}}},
  \bibinfo{author}{\bibfnamefont{M.}~\bibnamefont{{Dickinson}}},
  \bibinfo{author}{\bibfnamefont{H.~C.} \bibnamefont{{Ferguson}}},
  \bibinfo{author}{\bibfnamefont{T.~A.} \bibnamefont{{Hutchison}}},
  \bibinfo{author}{\bibfnamefont{J.~S.} \bibnamefont{{Kartaltepe}}},
  \bibinfo{author}{\bibfnamefont{R.~L.} \bibnamefont{{Larson}}},
  \bibinfo{author}{\bibfnamefont{R.~C.} \bibnamefont{{Simons}}},
  \bibinfo{author}{\bibfnamefont{C.}~\bibnamefont{{Papovich}}},
  \bibnamefont{et~al.}, \bibinfo{journal}{arXiv e-prints}
  \bibinfo{eid}{arXiv:2304.05385} (\bibinfo{year}{2023}), \eprint{2304.05385}.

\bibitem[{\citenamefont{Cai et~al.}(2023)\citenamefont{Cai, Tang, Mo, Yan,
  Chen, Ma, Wang, Luo, Easson, and Marciano}}]{Cai:2023ptf}
\bibinfo{author}{\bibfnamefont{Y.-F.} \bibnamefont{Cai}},
  \bibinfo{author}{\bibfnamefont{C.}~\bibnamefont{Tang}},
  \bibinfo{author}{\bibfnamefont{G.}~\bibnamefont{Mo}},
  \bibinfo{author}{\bibfnamefont{S.}~\bibnamefont{Yan}},
  \bibinfo{author}{\bibfnamefont{C.}~\bibnamefont{Chen}},
  \bibinfo{author}{\bibfnamefont{X.}~\bibnamefont{Ma}},
  \bibinfo{author}{\bibfnamefont{B.}~\bibnamefont{Wang}},
  \bibinfo{author}{\bibfnamefont{W.}~\bibnamefont{Luo}},
  \bibinfo{author}{\bibfnamefont{D.}~\bibnamefont{Easson}}, \bibnamefont{and}
  \bibinfo{author}{\bibfnamefont{A.}~\bibnamefont{Marciano}}
  (\bibinfo{year}{2023}), \eprint{2301.09403}.

\bibitem[{\citenamefont{Kinney}(2005)}]{Kinney:2005vj}
\bibinfo{author}{\bibfnamefont{W.~H.} \bibnamefont{Kinney}},
  \bibinfo{journal}{Phys. Rev. D} \textbf{\bibinfo{volume}{72}},
  \bibinfo{pages}{023515} (\bibinfo{year}{2005}), \eprint{gr-qc/0503017}.

\bibitem[{\citenamefont{Martin et~al.}(2013)\citenamefont{Martin, Motohashi,
  and Suyama}}]{Martin:2012pe}
\bibinfo{author}{\bibfnamefont{J.}~\bibnamefont{Martin}},
  \bibinfo{author}{\bibfnamefont{H.}~\bibnamefont{Motohashi}},
  \bibnamefont{and} \bibinfo{author}{\bibfnamefont{T.}~\bibnamefont{Suyama}},
  \bibinfo{journal}{Phys. Rev. D} \textbf{\bibinfo{volume}{87}},
  \bibinfo{pages}{023514} (\bibinfo{year}{2013}), \eprint{1211.0083}.

\bibitem[{\citenamefont{Garcia-Bellido and
  Ruiz~Morales}(2017)}]{Garcia-Bellido:2017mdw}
\bibinfo{author}{\bibfnamefont{J.}~\bibnamefont{Garcia-Bellido}}
  \bibnamefont{and}
  \bibinfo{author}{\bibfnamefont{E.}~\bibnamefont{Ruiz~Morales}},
  \bibinfo{journal}{Phys. Dark Univ.} \textbf{\bibinfo{volume}{18}},
  \bibinfo{pages}{47} (\bibinfo{year}{2017}), \eprint{1702.03901}.

\bibitem[{\citenamefont{Pi and Sasaki}(2023)}]{Pi:2021dft}
\bibinfo{author}{\bibfnamefont{S.}~\bibnamefont{Pi}} \bibnamefont{and}
  \bibinfo{author}{\bibfnamefont{M.}~\bibnamefont{Sasaki}},
  \bibinfo{journal}{Phys. Rev. D} \textbf{\bibinfo{volume}{108}},
  \bibinfo{pages}{L101301} (\bibinfo{year}{2023}), \eprint{2112.12680}.

\bibitem[{\citenamefont{Cai et~al.}(2018)\citenamefont{Cai, Tong, Wang, and
  Yan}}]{Cai:2018tuh}
\bibinfo{author}{\bibfnamefont{Y.-F.} \bibnamefont{Cai}},
  \bibinfo{author}{\bibfnamefont{X.}~\bibnamefont{Tong}},
  \bibinfo{author}{\bibfnamefont{D.-G.} \bibnamefont{Wang}}, \bibnamefont{and}
  \bibinfo{author}{\bibfnamefont{S.-F.} \bibnamefont{Yan}},
  \bibinfo{journal}{Phys. Rev. Lett.} \textbf{\bibinfo{volume}{121}},
  \bibinfo{pages}{081306} (\bibinfo{year}{2018}), \eprint{1805.03639}.

\bibitem[{\citenamefont{Cai et~al.}(2019)\citenamefont{Cai, Chen, Tong, Wang,
  and Yan}}]{Cai:2019jah}
\bibinfo{author}{\bibfnamefont{Y.-F.} \bibnamefont{Cai}},
  \bibinfo{author}{\bibfnamefont{C.}~\bibnamefont{Chen}},
  \bibinfo{author}{\bibfnamefont{X.}~\bibnamefont{Tong}},
  \bibinfo{author}{\bibfnamefont{D.-G.} \bibnamefont{Wang}}, \bibnamefont{and}
  \bibinfo{author}{\bibfnamefont{S.-F.} \bibnamefont{Yan}},
  \bibinfo{journal}{Phys. Rev. D} \textbf{\bibinfo{volume}{100}},
  \bibinfo{pages}{043518} (\bibinfo{year}{2019}), \eprint{1902.08187}.

\bibitem[{\citenamefont{Zhou et~al.}(2020)\citenamefont{Zhou, Jiang, Cai,
  Sasaki, and Pi}}]{Zhou:2020kkf}
\bibinfo{author}{\bibfnamefont{Z.}~\bibnamefont{Zhou}},
  \bibinfo{author}{\bibfnamefont{J.}~\bibnamefont{Jiang}},
  \bibinfo{author}{\bibfnamefont{Y.-F.} \bibnamefont{Cai}},
  \bibinfo{author}{\bibfnamefont{M.}~\bibnamefont{Sasaki}}, \bibnamefont{and}
  \bibinfo{author}{\bibfnamefont{S.}~\bibnamefont{Pi}}, \bibinfo{journal}{Phys.
  Rev. D} \textbf{\bibinfo{volume}{102}}, \bibinfo{pages}{103527}
  (\bibinfo{year}{2020}), \eprint{2010.03537}.

\bibitem[{\citenamefont{Cai et~al.}(2021)\citenamefont{Cai, Jiang, Sasaki,
  Vardanyan, and Zhou}}]{Cai:2021yvq}
\bibinfo{author}{\bibfnamefont{Y.-F.} \bibnamefont{Cai}},
  \bibinfo{author}{\bibfnamefont{J.}~\bibnamefont{Jiang}},
  \bibinfo{author}{\bibfnamefont{M.}~\bibnamefont{Sasaki}},
  \bibinfo{author}{\bibfnamefont{V.}~\bibnamefont{Vardanyan}},
  \bibnamefont{and} \bibinfo{author}{\bibfnamefont{Z.}~\bibnamefont{Zhou}},
  \bibinfo{journal}{Phys. Rev. Lett.} \textbf{\bibinfo{volume}{127}},
  \bibinfo{pages}{251301} (\bibinfo{year}{2021}), \eprint{2105.12554}.

\bibitem[{\citenamefont{Carr and Kuhnel}(2020)}]{Carr:2020xqk}
\bibinfo{author}{\bibfnamefont{B.}~\bibnamefont{Carr}} \bibnamefont{and}
  \bibinfo{author}{\bibfnamefont{F.}~\bibnamefont{Kuhnel}},
  \bibinfo{journal}{Ann. Rev. Nucl. Part. Sci.} \textbf{\bibinfo{volume}{70}},
  \bibinfo{pages}{355} (\bibinfo{year}{2020}), \eprint{2006.02838}.

\bibitem[{\citenamefont{Thrane and Talbot}(2019)}]{Thrane:2018qnx}
\bibinfo{author}{\bibfnamefont{E.}~\bibnamefont{Thrane}} \bibnamefont{and}
  \bibinfo{author}{\bibfnamefont{C.}~\bibnamefont{Talbot}},
  \bibinfo{journal}{Publ. Astron. Soc. Austral.} \textbf{\bibinfo{volume}{36}},
  \bibinfo{pages}{e010} (\bibinfo{year}{2019}), \bibinfo{note}{[Erratum:
  Publ.Astron.Soc.Austral. 37, e036 (2020)]}, \eprint{1809.02293}.

\bibitem[{\citenamefont{Akaike}(1974)}]{Akaike1974ANL}
\bibinfo{author}{\bibfnamefont{H.}~\bibnamefont{Akaike}},
  \bibinfo{journal}{IEEE Transactions on Automatic Control}
  \textbf{\bibinfo{volume}{19}}, \bibinfo{pages}{716} (\bibinfo{year}{1974}).

\bibitem[{\citenamefont{Wang et~al.}(2023{\natexlab{b}})}]{WFST:2023voz}
\bibinfo{author}{\bibfnamefont{T.}~\bibnamefont{Wang}} \bibnamefont{et~al.}
  (\bibinfo{collaboration}{WFST}), \bibinfo{journal}{Sci. China Phys. Mech.
  Astron.} \textbf{\bibinfo{volume}{66}}, \bibinfo{pages}{109512}
  (\bibinfo{year}{2023}{\natexlab{b}}), \eprint{2306.07590}.

\bibitem[{\citenamefont{Ziparo et~al.}(2022)\citenamefont{Ziparo, Gallerani,
  Ferrara, and Vito}}]{Ziparo:2022fnc}
\bibinfo{author}{\bibfnamefont{F.}~\bibnamefont{Ziparo}},
  \bibinfo{author}{\bibfnamefont{S.}~\bibnamefont{Gallerani}},
  \bibinfo{author}{\bibfnamefont{A.}~\bibnamefont{Ferrara}}, \bibnamefont{and}
  \bibinfo{author}{\bibfnamefont{F.}~\bibnamefont{Vito}},
  \bibinfo{journal}{Mon. Not. Roy. Astron. Soc.}
  \textbf{\bibinfo{volume}{517}}, \bibinfo{pages}{1086} (\bibinfo{year}{2022}),
  \eprint{2209.09907}.

\end{thebibliography}

\section*{Appendix: The Posterior Distribution of PBH Population}\label{appendix}
\begin{figure*}[htbp]
\includegraphics[width=0.34\textwidth]{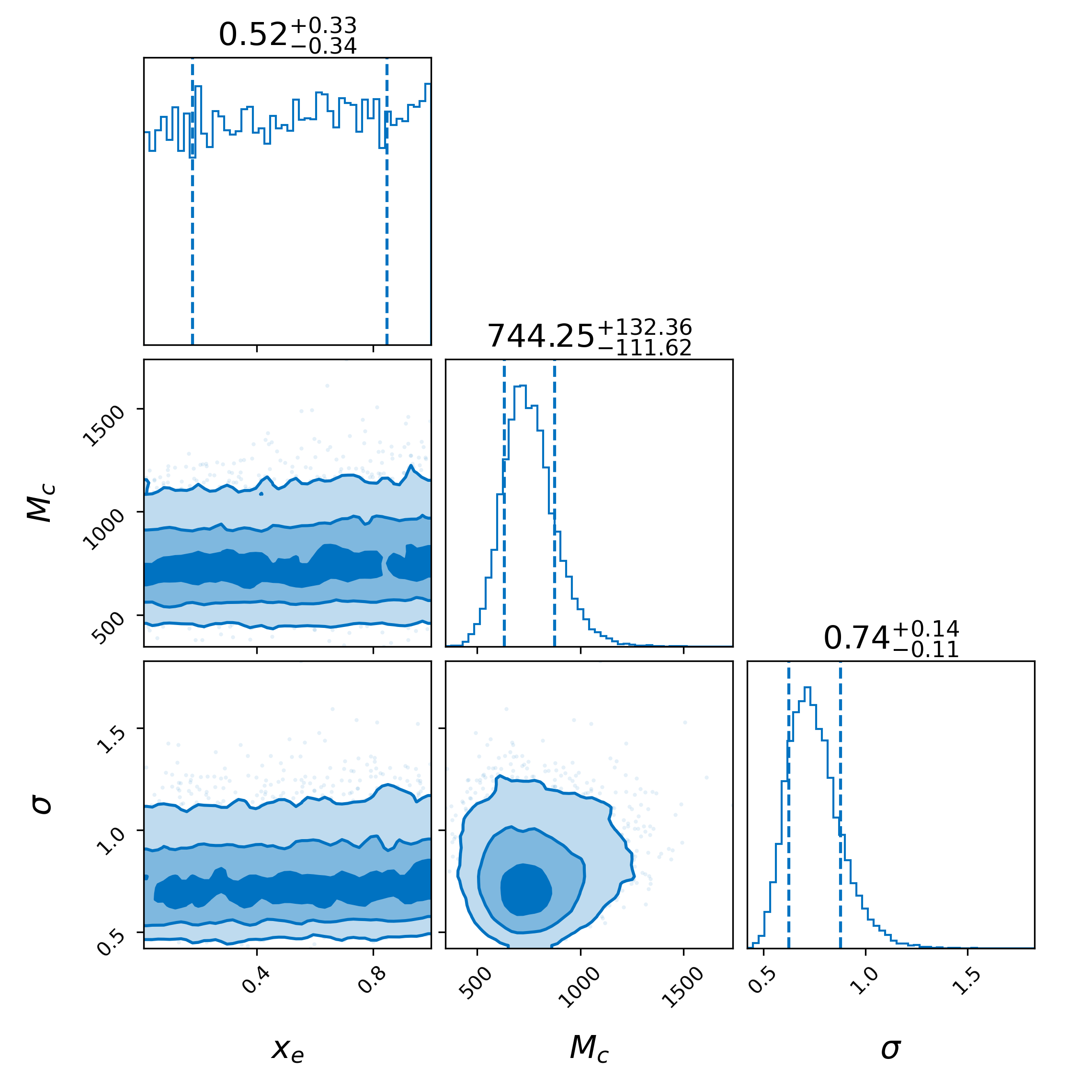}
\includegraphics[width=0.34\textwidth]{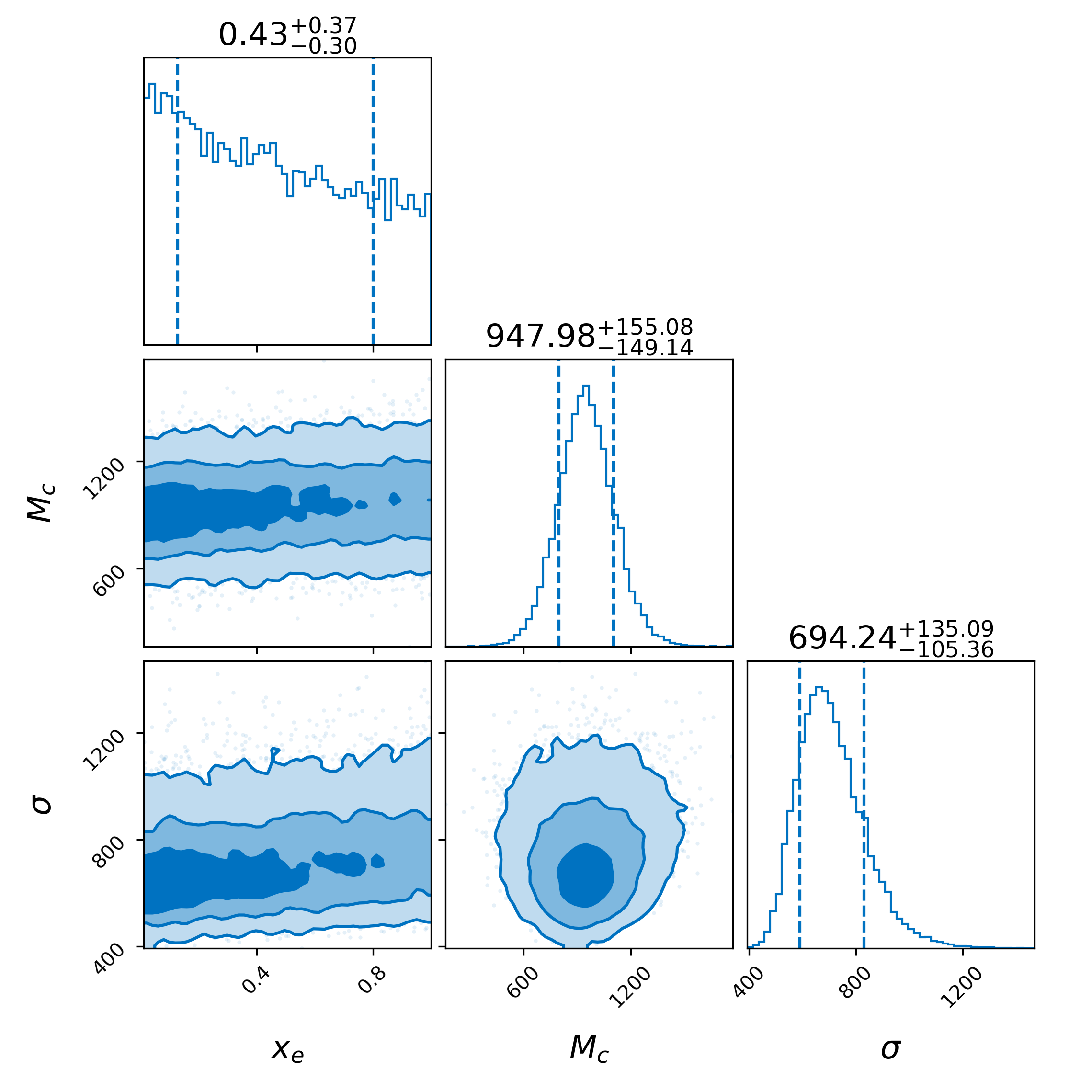}
\includegraphics[width=0.28\textwidth]{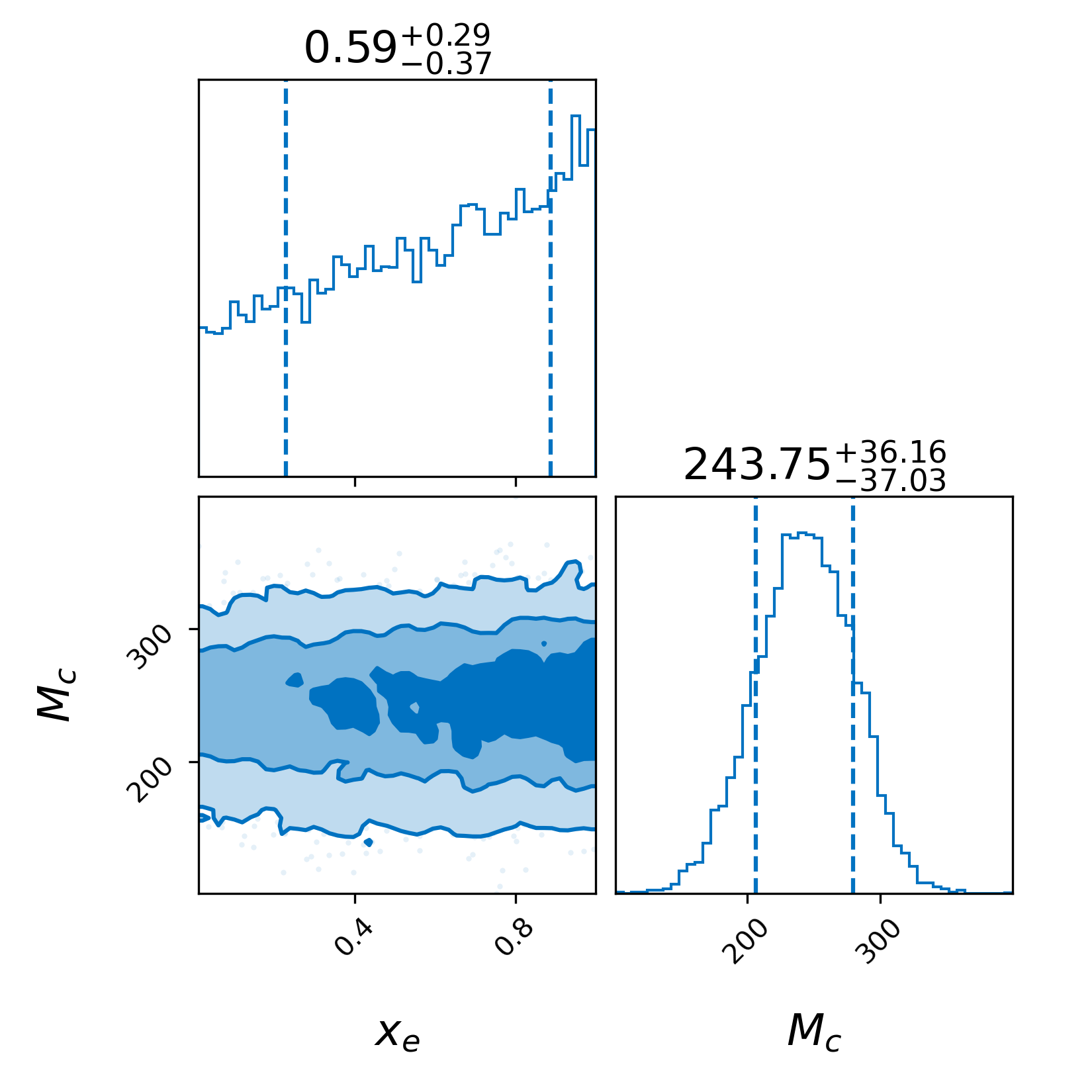}
\includegraphics[width=0.4\textwidth]{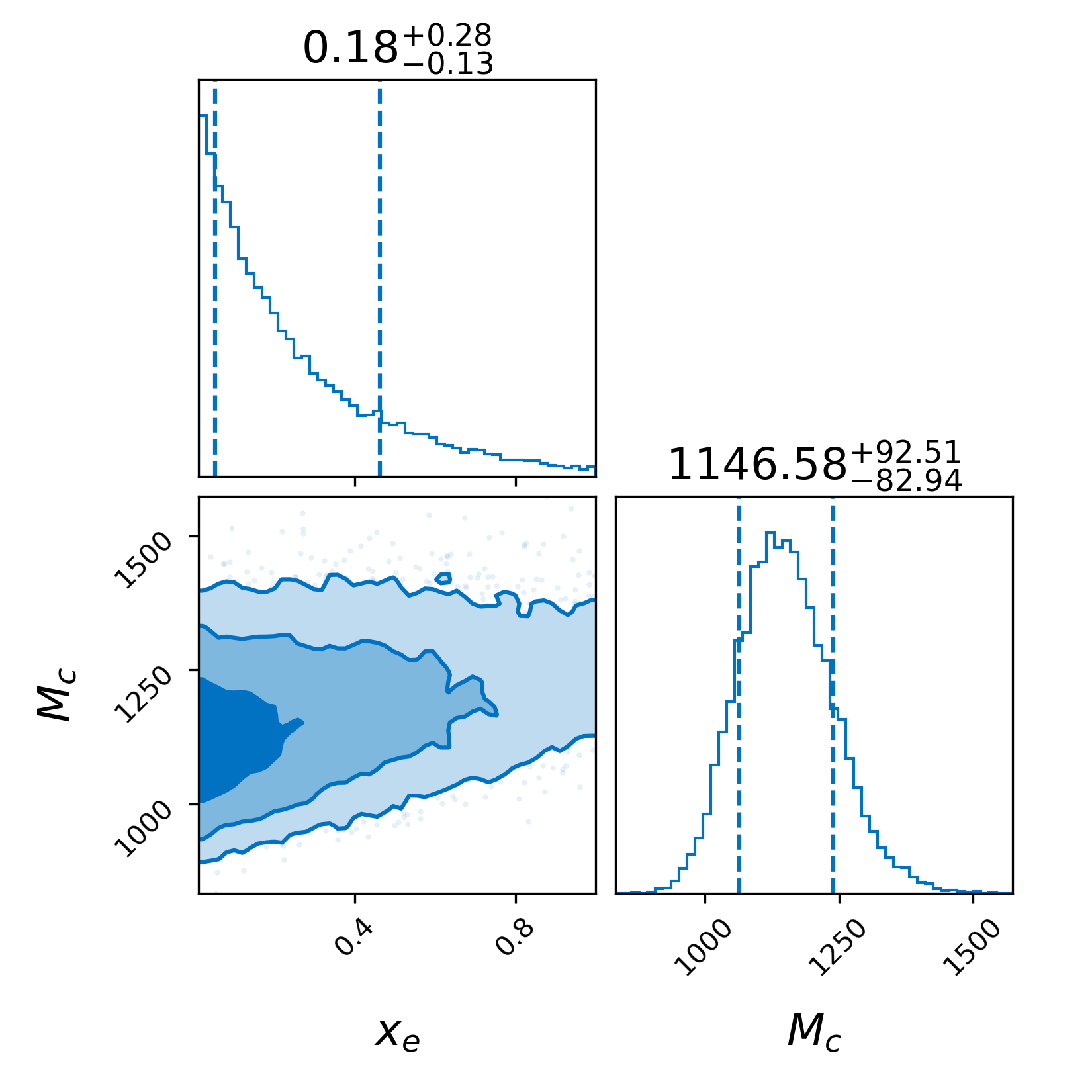}
\includegraphics[width=0.55\textwidth]{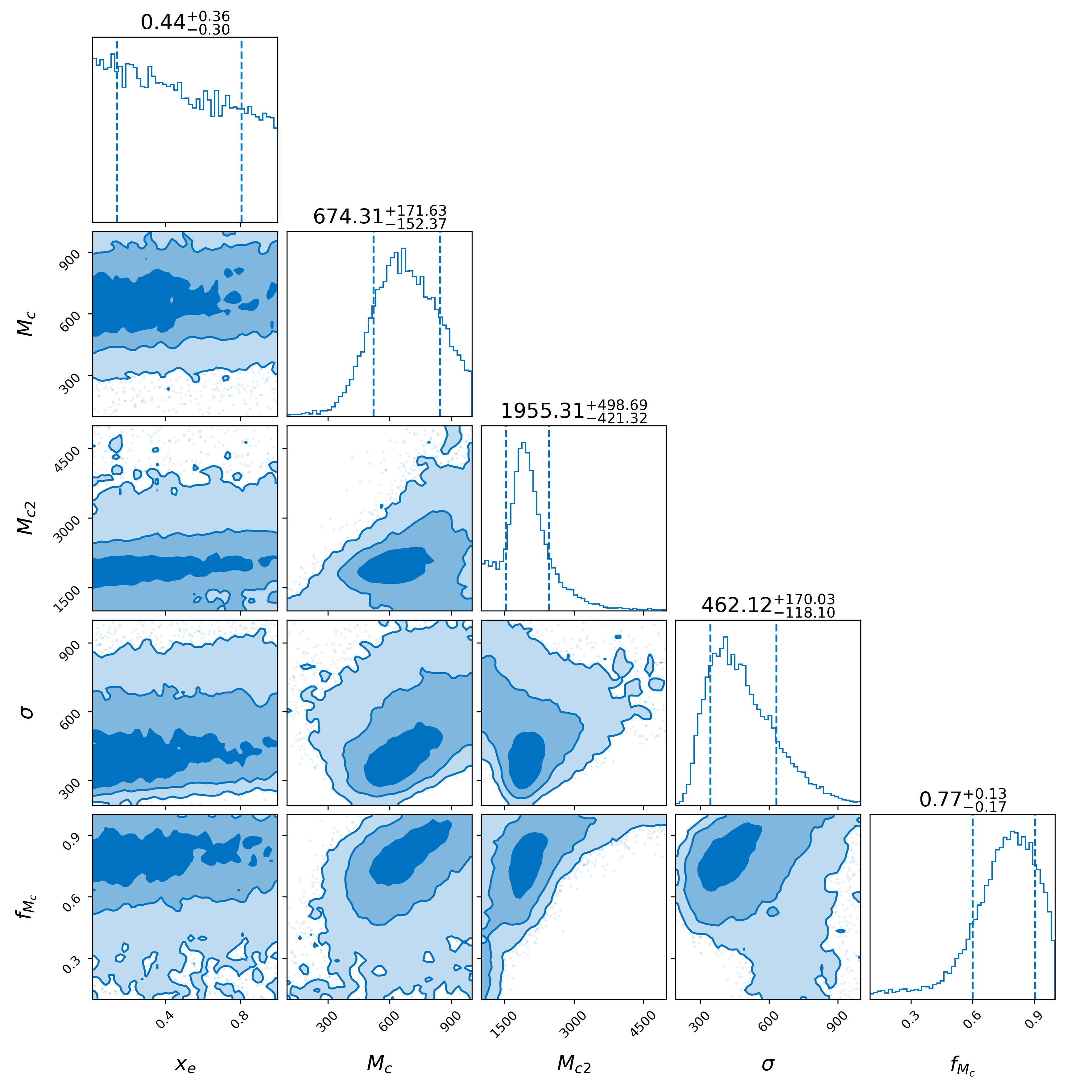}
\caption{The posterior distributions of individual parameters for five different population models. The upper panels from left to right panels show the results for the Lognormal, Gaussian and Powerlaw models, respectively. While the lower left and right panels refer to the Critical and Multipeak models. All models include the ionization fraction of the cosmic gas x$_e$, and we use the parameters listed in Table~\ref{evidence_table} with Uniform priors. The 68\% confidence intervals for each parameter are reported above each column, corresponding to the models shown in the inset.}
\label{five_corner}
\end{figure*}

The posteriors of the hyper-parameters describing the initial mass distributions of PBHs (Logormal/Gaussian/Powerlaw/Critical/Multipeak), as introduced in the main text, are displayed in Figure~\ref{five_corner}, respectively.

\begin{center}
\begin{table*}[htbp]
\setlength{\tabcolsep}{.1em}
\begin{tabular}{lccccc}
\hline
 Models & x$_e$ & $M_c$ &  $\sigma$ & $M_{c2}$ & $f_{M_c}$ \\
\hline
Lognormal & $\mathcal{U}(0.01,1.0)$ & $\mathcal{U}(10, 2500)$ &  $\mathcal{U}(0.01,5.0)$ & - & - \\
\hline
Powerlaw & $\mathcal{U}(1.0,10.0)$ & $\mathcal{U}(0.1,10.0)$ &  - & - & -  \\
\hline
Multipeak & $\mathcal{U}(0.01,1.0)$ & $\mathcal{U}(100,1000)$ & $\mathcal{U}(10,1000)$ & $\mathcal{U}(1000,5000)$ & $\mathcal{U}(0.1,1.0)$ \\
\hline
Gaussian & $\mathcal{U}(0.01,1.0)$ & $\mathcal{U}(10,10000)$ & $\mathcal{U}(10,3000)$ & - & - \\
\hline
Critical & $\mathcal{U}(0.01,1.0)$ & $\mathcal{U}(500,5000)$ &  - & - & - \\
\hline
\end{tabular}
\caption{The priors of the parameters in mass functions of PBH. The first column lists the names of the models, followed by their uniform distribution. }
\label{prior}
\end{table*}
\end{center}

\end{document}